\documentstyle[prl,preprint,aps,psfig]{revtex}
\begin{document}
\draft
\title{Excitons of Composite Fermions}
\author{R.K. Kamilla, X.G. Wu, and J.K. Jain}
\address{Department of Physics, State University of New York
at Stony Brook, Stony Brook, New York 11794-3800}
\date{\today}
\maketitle
\begin{abstract}

The low-energy excitations of filled Landau levels (LL's) of 
electrons involve promotion of a single electron from the topmost 
filled LL to the lowest empty LL. These are called excitons or
collective modes. The incompressible fractional quantum Hall states
are understood as filled LL's of composite fermions, and 
the low-energy neutral excitations are excitons of composite
fermions. New techniques are developed to study large systems,
which provide detailed information about the dispersions of 
the composite fermion excitons. In particular, it is found that the
interaction energy of the exciton is well described by the 
`unprojected' composite fermion theory.
\footnotetext{Accepted for publication in Physical Review B (1996).}
\end{abstract}


\section{Introduction}

The quantum Hall effect (QHE) is one of the most 
fascinating discoveries in physics in recent years. It
concerns electron transport in two dimensional systems in the
presence of a high transverse magnetic field. It has been found that
there are plateaus on which the Hall resistance is quantized \cite
{KvonK} at
\begin{equation}
R_{H}={\displaystyle{h \over fe^2}}
\end{equation}
where $f$ is either an integer or a simple fraction. For historical
reasons, the observation of integer values of $f$ is called the
integer QHE (IQHE) and of fractional values of $f$ is called the 
fractional QHE (FQHE). The longitudinal resistance in the plateau 
region is exponentially small, vanishing in the limit of temperature
$T\rightarrow 0$.

Not long after the discovery of the FQHE, Laughlin \cite {Laughlin1} 
proposed the following wave function for the ground state at  
$f={\displaystyle {1 \over (2m+1)}}$:
\begin{equation}
\psi_{{1/(2m+1)}}= \prod_{j<k}(z_{j}-z_{k})^{2m+1}
\exp[-\frac{1}{4}\sum_{i}|z_{i}|^2]
\end{equation}
where $z_{j}=x_{j}+iy_{j}$ denotes the coordinates
($x_{j}\;,y_{j}$)
of the $j$th electron, and the lengths are expressed in
unit of the magnetic length $l_{0}=\sqrt{\displaystyle{\hbar c \over eB}}$.
This wave function was studied for small systems and found to be quite
accurate. Laughlin also wrote trial wave functions for the 
charged excitations (quasielectrons and quasiholes, collectively
referred to as quasiparticles). 

Subsequently,  
Girvin, MacDonald and Platzman \cite {GMP} (GMP) employed a single mode
approximation (SMA) to investigate the neutral excitation of the 
Laughlin state, called the collective mode (CM) or exciton. 
The SMA takes the trial wave function of the 
collective excitation to be
\begin{equation}
\chi_{_{k}}^{SMA}={\cal P}\rho_{_{k}}\psi_{1/(2m+1)}
\label{SMAwf}
\end{equation}
where ${\cal P}$ is the
lowest-Landau-level (LLL) projection operator and 
\begin{equation}
\rho_{_{k}}=\sum_{j=1}^{N}e^{i{\bf k}\cdot {\bf r}_{j}}
\end{equation}
is the density wave operator with wave vector
$k$, which is the Fourier transform of the density
\begin{equation}
\rho({\bf r})=\sum_{j=1}^{N}\;\delta({\bf r}-{\bf r}_{j})\;,
\end{equation}
for $N$ electrons.
The notation $\chi$ will be used exclusively for the wave function of
the collective mode or exciton
in this article; the superscript will denote which
scheme is used, and the subscript will show the quantum
number. GMP evaluated the energy of this wave function,
given by
\begin{equation}
\frac{<\chi_{_{k}}^{SMA}|H-E_{0}|\chi_{_{k}}^{SMA}>}
{<\chi_{_{k}}^{SMA}|\chi_{_{k}}^{SMA}>}\;,
\end{equation}
as a function of $k$.  In finite system studies,
the SMA was found to work well in an intermediate range of wave vectors 
near a minimum in the dispersion, termed the ``roton" 
minimum, in analogy with Feynman's theory of superfluid $^4$He 
\cite {Feynman}.  At wave vectors beyond the minimum, however, 
the SMA was not satisfactory. For large $k$, the
energy of the true collective mode should be roughly independent of $k$, 
since it contains a far separated
quasielectron-quasihole pair. The energy of $\chi_{k}^{SMA}$,
however, continues to grow with $k$. Laughlin attempted 
an alternative description of the neutral collective mode  
in terms of a quasielectron-quasihole pair excitation 
\cite {Laugha}. This gave a reasonable description at large wave 
vectors, but did not produce the roton minimum. Thus, it was not 
possible to describe the full dispersion of the neutral
excitation of $f=1/(2m+1)$ in a single theoretical framework.

Much less was known about the collective modes of the other 
FQHE states. The SMA can be generalized to the other fractions as
\begin{equation}
\chi_{_{k}}^{SMA}={\cal P}\rho_{_{k}}\psi_{n/(2mn+1)}\;,
\end{equation}
where $\psi_{n/(2mn+1)}$ is the incompressible ground state at
$f={\displaystyle {n \over (2mn+1)}}$. Since no good microscopic 
ansatz existed for $\psi_{n/(2mn+1)}$, the SMA  
was tested in finite system studies, with the exact numerical ground
state used for $\psi_{n/(2mn+1)}$ \cite {Su}.
It, however, did not give a satisfactory description  of the 
collective excitations \cite {Su}. 
Finite system exact diagonalization 
studies \cite {He} also failed to provide a 
satisfactory overall picture for the collective excitations at general
fractions. 
The results showed a significant amount of finite size effects, making
an extrapolation to the thermodynamic limit difficult.

In the last few years, there has been a resurgence of
interest in this issue for two reasons. First, significant  
progress has been made on the experimental front.
Pinczuk {\em et al.} \cite {Pinczuk1} have measured the positions of the
maxima and minima in the collective modes of the 
IQHE states, and recently also their  
dispersion in modulated density samples \cite {Pinczuk2}. 
Further, Raman scattering \cite {Pinczuk} 
and phonon absorption \cite {Mellor} experiments have 
reported observation of the collective modes in the 
FQHE regime.  

Secondly, there now exists a new theoretical framework, 
called the composite fermion (CF) theory \cite {Jain}, for 
describing all FQHE on an equal footing. The strongly correlated 
electrons are mapped on to weakly correlated 
composite fermions, which are electrons `dressed' with $2m$ vortices of the
many particle wave function. They may be loosely thought of as
electrons carrying $2m$ flux quanta. The composite fermions experience a 
reduced effective magnetic field, and form energy levels, called 
CF-LL's or quasi-LL's, which are similar to the LL's of electrons 
in this {\em reduced} magnetic field.  The FQHE of electrons is 
understood as the IQHE of composite fermions, and the FQHE ground
state at $f=n/(2mn+1)$ is interpreted as the state containing $n$  
filled CF-LL's. The Laughlin wave function, in particular, 
is interpreted as one filled CF-LL. 
This description of the incompressible FQHE states 
suggests that the low-energy neutral 
excitations can be obtained by promoting a single
composite fermion to the next higher CF-LL (Fig.~1), 
i.e., they are excitons of composite fermions.
Their wave functions can be constructed in analogy to the
wave functions of the excitons of electrons for IQHE states,
and exact-diagonalization studies for systems
containing up to 9 electrons have shown that they 
provide an extremely good quantitative description of the 
low-energy excitations of the FQHE
states at 1/3, 2/5, and 3/7 for all values of $k$ accessible in these
finite size systems \cite {Dev,Wu}. 

The excited composite fermion will be called a quasielectron of the
FQHE ground state, and the hole left behind a quasihole \cite
{Goldhaber}.
The essential difference between the SMA and the CF wave functions is
that the SMA creates an {\em electron-hole} pair, whereas the CF wave 
function creates a {\em quasielectron-quasihole} pair.
We will see below that the SMA and the CF approaches are in general
unrelated; the only exception is for the exciton at  
${\displaystyle {1 \over (2m+1)}}$, for which the two 
become equivalent in the limit $k\rightarrow 0$.

Despite the confirmation of the validity of the CF scheme in small
system studies, it has not been possible to study large systems yet.
The main difficulty is that the CF wave functions 
have a small amount of mixing with higher LL's to begin
with, and must be projected on to the lowest LL to obtain good
quantitative estimates in the large $B$ limit.
The projection operator is quite complicated  
and no general computational
schemes exist for dealing with it for large systems.
In the present study, we  develop two
techniques for computing the properties of CF excitons for large
systems. The first one computes the energy of the 
LLL-projected wave function of the CF exciton of the 
$\displaystyle{1 \over (2m+1)}$ 
state by generalizing a method developed 
by Bonesteel \cite {Bonesteel}. This method, however, is inapplicable 
to other FQHE states. In the second one, we show that the `unprojected' 
wave functions themselves can be used to a great extent 
for the study of the exciton dispersion. We have computed 
the (self) energy of a single quasielectron or quasihole in the unprojected 
scheme, and found it to be off by a factor of two to three \cite
{unpub}. However, 
the interaction between the quasielectron and
quasihole participating in the exciton is found to be well described
by the unprojected theory, except when they are very close. Therefore,
the unprojected theory gives the exciton dispersion reasonably well up
to an additive constant correcting for error in the self energies of
the quasielectron and quasihole. We study large systems within the framework
of the unprojected CF theory, and make qualitatively new and detailed 
predictions for various collective mode dispersions, which  
exhibit a rich structure with several minima.  The extrema
in the dispersion are of experimental relevance, since the CM 
density of states has 
peaks at the corresponding energies, which, as a result 
of a disorder-induced breakdown of the wave vector conservation,
become observable in inelastic light scattering experiments 
\cite {Pinczuk1}. 

The analogy to the IQHE excitons is helpful in
explaining the number of minima and their positions. The relatively deep 
minima in the CM dispersion of the 
$f={\displaystyle {n \over (2mn+1)}}$ FQHE 
state occur at approximately the same wave vector as those in the CM
dispersion of the $f=n$ IQHE state. The latter originate from an
interplay between the density profiles of the quasielectron in the
$(n+1)$th LL and the
quasihole in the $n$th LL as they approach one another. Our study thus
indicates that the wave function profile of the quasielectrons in the
$f={\displaystyle {n \over (2mn+1)}}$ FQHE state is similar to that 
in the $f=n$ IQHE
state, further deepening the analogy between the FQHE and the
IQHE implied by the CF theory. 

The dispersions for the collective modes have also been calculated 
in a Chern-Simons (CS) field theoretical scheme in several recent 
papers. Here, the composite fermions are modeled  as electrons carrying 
(gauge) flux quanta \cite {Jain,Lopez1}, which simulate vortices.
The qualitative physical picture of composite fermions is obtained at a
mean-field level, and a perturbation theory is carried out at the
level of random-phase-approximation (RPA) \cite {Lopez2}. Although the 
trial-wave-function and the CS schemes are both attempting to 
describe the same physics from different starting points, 
a precise relationship between them is not known at the moment.
We will point out some similarities and differences between the results
of the two methods in the specific context of the collective modes.

The plan of the paper is as follows.
In Section II, we write the wave function of the CF exciton,
and show that, for $f=1/(2m+1)$, it is equal to SMA wave function 
in the limit
of small wave vectors. Section III gives the results of our study of
the projected CF excitons, both in small systems, where a comparison with
the exact results is possible, and in large systems, where Monte Carlo
techniques \cite {Binder,Ceperley} allow the 
computation of its dispersion for the 1/3
state. In Section IV, we study the interaction energy of the
unprojected CF-exciton wave functions. It is shown that it provides a 
reasonably good quantitative description of the {\em interaction}
between the  quasielectron and the quasihole. 
Section V considers 
the analogy between the excitons of composite fermions, relevant for
the FQHE, and those of electrons, relevant for 
the IQHE. This provides insight into the
results of the previous sections, and also sheds light on the
origin of the minima in the FQHE exciton dispersions. Section VI
investigates the kinetic energies of the unprojected wave functions. 
It is shown to be less than the cyclotron energy for finite $k$, 
approaching the cyclotron energy as $k\rightarrow 0$. This is 
reminiscent of the Kohn's theorem, a weaker form of which is given 
for a class of wave functions which are not eigenstates of any known 
Hamiltonian. The paper is concluded in Section VII.
A short report of parts of this paper has been published elesewhere
\cite {KWJ}.

For future reference, here is a list of notations we will be using 
below for various wave functions and energies.

\noindent
$\Phi_n$: Slater determinant wave function of $n$ filled LL's of
electrons.

\noindent
$\Phi_{n}^{CF}$ ($\Phi_{n}^{UP-CF}$): Projected (unprojected) wave
function of $n$ filled LL's of composite fermions; this corresponds to
the ground state of electrons at $f=n/(2mn+1)$. The superscript `UP'
denotes `unprojected'. 

\noindent
$\chi^{CF}$ ($\chi^{UP-CF}$):  Projected (unprojected) wave
function of the CF exciton.

\noindent
$\Delta V_{ex}$: Exact energy of the exciton (measured relative to the
exact ground state) from exact diagonalization in the lowest Landau
level.

\noindent
$\Delta V_p$: Interaction energy of $\chi^{CF}$ relative to
$\Phi^{CF}$.

\noindent
$\Delta V$ ($\Delta K$): Interaction (kinetic) 
energy of $\chi^{UP-CF}$ relative to $\Phi^{UP-CF}$.

\section{composite fermion excitons}

The basic principle of the CF theory \cite {Jain} is that, in a 
range of filling factors, the electrons in the lowest 
LL find it energetically favorable to capture an even number of
vortices of the many particle wave function. The bound state of an
electron and the vortices behaves as a particle, called 
composite fermion. The vortices produce phases as the composite
fermions move around, which partly cancel the Aharonov-Bohm phases  
originating from the external magnetic field, and, as a result, 
the composite fermions experience an effective magnetic field 
given by 
\begin{equation}
B^*=B-2m\rho\phi_{0}\;,
\end{equation}
where $B$ is the external field, $2m$ is the number of vortices 
carried by composite fermions, $\phi_{0}=hc / e$ is the flux quantum, 
and $\rho$ is
the electron (or CF) density. The residual interaction between the
composite fermions is weak, and the strongly correlated liquid of
electrons maps into a  weakly interacting gas of composite fermions.
An effective single-particle description of the electron state thus 
becomes possible in terms of composite fermions.  The energy levels 
of composite fermions are analogous to the LL's of 
non-interacting electrons in this {\em weaker} magnetic field, 
and are called quasi- or CF-LL's.  Defining the CF filling factor 
as $\nu^*=\rho \phi_{0} / B^*$, in analogy to the
electron filling factor $\nu=\rho \phi_{0}/B$, the above equation 
can also be expressed as 
\begin{equation}
\nu={\nu^* \over (2m\nu^*+1)}\;.
\end{equation}
The IQHE of composite fermions at $\nu^*=n$ manifests as the
FQHE of electrons at $\displaystyle {\nu={n \over (2mn+ 1)}}\;.$

The trial wave functions for the CF states are constructed as 
follows. For
simplicity, we confine our discussion to the special filling factors 
$\displaystyle {\nu={n \over (2mn+1)}}$,  
which corresponds to the CF filling factor $\nu^*=n$.
Let us denote the ground state of non-interacting electrons at
$\nu^*=n$ by $\Phi_{n}$. The wave function of $n$
filled CF-LL's is obtained by attaching $2m$ vortices to each 
electron in the state $\Phi_{n}$, which amounts to a multiplication 
by the Jastrow factor 
\begin{equation}
D^m \equiv \prod_{j<k}(z_{j}-z_{k})^{2m}\;,
\end{equation}
which produces 
\begin{equation}
\Phi_{n}^{UP-CF}=\prod_{j<k}(z_{j}-z_{k})^{2m}\Phi_{n}\;.
\end{equation}
This wave function is not strictly in the lowest LL (although it is
largely so \cite {Trivedi}), and must be projected on to the LLL to
obtain a wave function appropriate to the limit $B\rightarrow \infty$:
\begin{equation}
\Phi_n^{CF}={\cal P}\Phi_{n}^{UP-CF}\;.
\end{equation}
$\Phi_n^{CF}$ describes the electron 
ground state at $\displaystyle {\nu={n \over (2mn+1)}}$.
It has been tested for
several FQHE states in small system exact calculations and found to 
be extremely accurate \cite {Dev,Wu}.

Armed with this analogy between the IQHE and the FQHE states, we 
are now in a position to write a wave function for the CF exciton (or
the CF collective mode).
It is obtained from the exciton of the $\nu=n$ IQHE state
precisely in the same manner as the FQHE ground state is obtained from
the $\nu=n$ IQHE ground state, i.e., by multiplication by the Jastrow
factor, followed by LLL projection:
\begin{equation}
\chi_{_{k}}^{CF}={\cal P}\chi_{_{k}}^{UP-CF}\;,
\label{cfcm}
\end{equation}
where
\begin{equation}
\chi_{_{k}}^{UP-CF}=
\prod_{j<k}(z_{j}-z_{k})^{2m}\;
\rho_{_{k}}^{n\rightarrow n+1} \Phi_{n}
\end{equation}
Here, $\rho_{_{k}}^{n\rightarrow n+1}\Phi_{n}$ is the exciton  
of the $\nu=n$ state.
In this wave function, a single
electron has been excited from the $n^{\textstyle {th}}$ LL 
of $\Phi_{n}$ to the $(n+1)^{\textstyle {th}}$ LL, creating an 
exciton whose kinetic energy is equal to 
the cyclotron energy at the reduced magnetic field,
$\displaystyle{\hbar\omega_{c}^*={\hbar\omega_{c} \over (2mn+1)}}$
\cite {Kallin}.  By analogy, $\chi^{CF}$ contains 
a single excited composite fermion in the $(n+1)^{\textstyle {th}}$
CF-LL and a hole left behind in the $n^{\textstyle {th}}$ CF-LL, as
shown schematically in Fig.~1.

The SMA wave function for the collective mode at $\nu=\displaystyle
{1 \over (2m+1)}$ has a CF-type interpretation, as it is 
obtained from $\rho_{_{k}}\Phi_{1}$ after  
multiplication by the Jastrow factor and then projecting the product
on to the lowest LL. We now show that the CF and the SMA wave 
functions become identical in the limit of $k\rightarrow 0$ for  
this mode.  This follows because, in general, in the limit $k\rightarrow 0$,
\begin{equation}
\rho_{_{k\rightarrow 0}}\Phi_{n}=
\rho_{_{k\rightarrow 0}}^{n\rightarrow n+1}\Phi_{n}\;.
\label{k0}
\end{equation}
To see this, we express $\rho_{_{k}}$
and $\rho_{_{k}}^{n\rightarrow n+1}$ in the second quantized form. 
Denote the single particle states by $|p,s>$, where $p$ is the wave
vector quantum number and $s=0,1,...$ is the LL index (note that the
topmost occupied LL of $\Phi_n$ is $s=n-1$). The 
explicit form of
the wave function in the Landau gauge is given by
\begin{equation}
\phi_{p,s}=(2\pi 2^s s!\sqrt{\pi})^{-1/2} \exp[ipy-\frac{1}{2}
(x+p)^2]\;H_{s}(x+p)\;\;,
\end{equation}
where $H_{s}$ is the Hermite polynomial, and the magnetic length has
been set equal to unity. In the second quantized notation, the density
operator is given by
\begin{equation}
\rho_{_{k}}=\sum_{p,p',s,s'}a^{\dagger}_{p's'}a_{ps}
<p's'|e^{iky}|ps>\;
\end{equation}
where the y-axis has been chosen parallel to the wave vector $k$. 
In the present case, we have $s<n$ and $s'\geq n$ since 
$\rho_{_{k}}$ is applied to a state with $n$ filled LL's.
Aside from an overall normalization factor, it is given by, for
$s'>s$,
\begin{equation}
\rho_{_{k}}=\sum_{s,s'} (2^{s+s'}s!s'!)^{-1/2}e^{-k^2/4}2^{-s}s!
k^{s'-s} L_{s}^{s'-s}(k^2/2)\ 
\int dp a^{\dagger}_{p+k,s'}a_{ps}\;. 
\label{rhoexp}
\end{equation}
The operator $\rho_{_{k}}^{n\rightarrow n+1}$ is given by the term with
$s=n-1$ and $s'=n$. Eq.~(\ref{k0}) follows in the limit $k\rightarrow
0$.  For fractions other than $\nu=1/(2m+1)$, 
$\chi^{SMA}$ does not have a CF-type
interpretation, and $\chi^{CF}$ and  $\chi^{SMA}$ are different at all $k$.

\section{Projected wave function}

This section investigates the projected CF wave function. 
The spherical geometry \cite {CNYang,book} is used in 
all our calculations, where $N$ electrons move on the surface of a
sphere under the influence of a radial magnetic field. 
The flux through the surface is $2q\phi_0$, where $2q$ is an integer.
The single particle eigenstates are called monopole harmonics,
$Y_{q,\ell,\ell_z}$, where $\ell=|q|,\;|q|+1,\;...$ is the single
particle angular momentum,
and $\ell_z=\pm \ell, \; \pm \ell-1,\;...$ is the z-component of
$\ell$. The total orbital angular momentum $L$ of the many body system 
is related to the wave vector
of the planar geometry by $\displaystyle{kl_{0}={L \over \sqrt{q}}}\;$ 
\cite {book}.  In the spherical geometry the energy
separation between the $n^{\textstyle{th}}$
and the $(n+1)^{\textstyle{th}}$ LL is given
by $\displaystyle {\left (1 + {n \over q}\right ) {\hbar eB \over
mc}}$, which reduces to the usual cyclotron energy
in the limit $q\rightarrow \infty$. We find it more convenient to 
use as our unit of the kinetic energy 
$\hbar\omega_c\equiv
\displaystyle{\left (1+{1 \over q}\right ){\hbar eB \over mc}}\;,$
the separation between the lowest two LL's.

If the highest occupied LL shell in $\Phi_{n}$
has angular momentum $\ell$, then the IQHE exciton has 
a single excited electron in the 
$(n+1)^{\textstyle{th}}$ LL, with  angular momentum $\ell+1$, 
and the hole in the $n^{\textstyle{th}}$ 
LL, with angular momentum $\ell$.  The Slater determinant 
basis states can be denoted by 
$$|\ell_z^{e},\ell_z^{h}>$$
where $\ell_z^{e}$ ($\ell_z^{h}$) is the z-component of the angular momentum
of the electron in the $(n+1)^{\textstyle {th}}$ LL 
(hole in the $n^{\textstyle{th}}$
LL). The allowed values of $L$ for the exciton  
are $L=1,\;2 \;...2\ell+1$, with 
precisely one multiplet at each $L$, containing $2L+1$ degenerate states. 
The $k= 0$ in the planar geometry corresponds to $L=1$ in the
spherical geometry, which is the smallest $L$ possible for the exciton.
With no loss of generality, we restrict our calculations below to 
the sector with $L_{z}=0$, with the
understanding that each state in this sector represents
$2L+1$ degenerate states of the full Hilbert space.
The exciton wave function $\rho_{_{L}}^{n\rightarrow n+1}\Phi_{n}$
is a {\em unique} linear superposition of the Slater 
determinant basis states, given by (with $L_{z}=0$)
\begin{equation}
\rho_{L}^{n\rightarrow n+1}\Phi_{n}
=\sum_{\ell_z=-\ell}^{\ell}
<\ell+1,\ell_z;\ell,-\ell_z|L,0> \;|\ell_z,-\ell_z>\;,
\end{equation}
The wave function of the CF exciton at the corresponding $L$, 
$\chi_{_{L}}^{CF}$, is obtained by multiplication by the Jastrow
factor, as in Eq.~(\ref{cfcm}). The Jastrow factor in the spherical
geometry is $\Phi_1^2$, where $\Phi_1$ is the wave function of the
lowest filled LL, multiplication by which does not change
the $L$ of the state. Clearly, $\chi_{_{L}}^{CF}$ 
{\em also does not contain any adjustable parameters}.

We have computed the exciton energy using the
projected CF wave function $\chi^{CF}$, denoted by $\Delta V_{p}$,
for finite systems.
It is measured relative to the projected CF ground state energy.
Our brute force projection
method (see Ref. \cite {Wu} for details)
allows us to carry out the projection for general states only
for up to 8-9 electrons, which severely limits the size of the systems for
which $\Delta V_{p}$ may be evaluated. 

Figs. 2 and ~3 give the actual Coulomb 
energies, $\Delta V_{ex}$,
obtained by an exact numerical diagonalization of the 
Hamiltonian, along with the energies of the
projected CF wave functions. First note that the range over which the
exciton extends in the exact diagonalization studies agrees 
with that predicted by the CF theory. (A state exists at $L=1$
for the IQHE exciton, but it is
annihilated by the projection operator while constructing the wave
function for the CF exciton.) $\Delta V_p$ is  
within a few percent of the exact energy $\Delta V_{ex}$.

For the exciton at $\nu=1/3$, we have obtained
$\Delta V_{p}$ for large systems using variational Monte Carlo, which
relies on the special feature of 
$\chi^{UP-CF}$ that it contains no
more than one electron in the second LL, and none at all in the 
higher LL's. This follows since the IQHE exciton wave function, 
$\rho_{_{k}}^{1\rightarrow 2} \Phi_{1}$, has only one electron in the
second LL, and multiplication by the Jastrow factor does not promote
electrons to higher LL's.
The projected wave function can then be written as \cite {Bonesteel}
\begin{equation}
\chi^{CF} \propto (K-\hbar\omega_{c}) \chi^{UP-CF}
\end{equation}
where,  we remind the 
reader that $\displaystyle{\hbar\omega_{c}=(1+{1 \over q})\hbar{eB \over mc}}$.
The kinetic energy operator $K$, in  standard
notation is given by
\begin{equation}
K = {\hbar \omega_c \over 2 q}\sum_{j=1}^N \left [\; -\; {1 \over
\sin\theta_j} \left ( {\partial \over \partial\theta_j} \left
(\sin\theta_j
{\partial \over \partial \theta_j}\right ) \right ) \; + \;
{1 \over \sin^{2}\theta_j}(q\cos\theta_j - m_j)^2 \right]
\; -\; {N \over 2}\hbar \omega_c
\end{equation}
where our choice of the gauge (or, `section' \cite {CNYang})
is defined by 
$i\partial_{\phi}Y_{q,\ell,\ell_z} \equiv (q - \ell_z)Y_{q,\ell,\ell_z}$,
and the kinetic energy is measured relative to the LLL.
Each Slater determinant is updated at every step of the Monte Carlo
following the method of Ref.~\cite {Ceperley}, which takes ${\cal
O}(N^2)$ operations.  The computation of the energy   
is rather involved and time consuming not only due to the presence of
derivatives in the expression of $K$, but also from the fact
that $\chi^{UP-CF}$ is a linear superposition of $\sim {\cal O}(N)$ 
Slater determinants, necessitating ${\cal O}(N^3)$ operations for
updating the determinants at each step. It should also be mentioned
that the computation of the interaction energy of $\chi^{CF}$ is much
more stable when $|\chi^{UP-CF}|^2$, rather than $|\chi^{CF}|^2$, is
chosen as the Monte Carlo weight function.

The CF-exciton energy is shown as a function of $L$ for 
a 20 electron system in Fig.~4.
Computation of each point takes approximately 5-6 days of computer
time on the IBM RISC 6000 workstation.
The $k\rightarrow 0$ ($L=1$) energy, $0.15e^2/\epsilon l_{0}$,
is consistent with the SMA prediction.  There is a deep minimum 
at the expected position of $kl_{0} \approx 1.4$. 
From Fig.~5, the estimate for the thermodynamic value of the
energy at the minimum is  $0.063(3)e^2/\epsilon l_{0}$,
which should be compared to the SMA prediction 
$0.078e^2/\epsilon l_{0}$ \cite {GMP}.
We note here that the same wave function for the ground state (the Laughlin 
wave function) is used in both the CF and the SMA calculations.

Two additional minima are clearly visible at $kl_{0}\approx 2.7$
and $kl_{0}\approx 3.5$.
Are they real? We believe so.  A direct confirmation of the genuineness
of the minimum at $kl_{0} \approx 2.7$ is 
seen in the nine-electron exact diagonalization calculation of Fano
{\em et al.} \cite {Fano}, reproduced in Fig.~6(c). (The exact
diagonalization systems are too small to see the third minimum.)
We note that there is no principle that
rules out the existence of more than one
minimum in the dispersion of the 1/3 exciton;  the structure 
here arises simply from an interplay between the structures
in the density profiles of thequasielectron  
and quasihole, as the distance between them is varied.
This will be discussed later in greater detail.

We note that the exciton energy predicted by the projected CF theory 
may not be very accurate in the limit $k\rightarrow 0$, as indicated 
by the following considerations. (i) The LLL
projection of the unprojected state becomes small and vanishes as
$k\rightarrow 0$ (i.e., at $L=1$), 
suggesting that the act of projection may not be as
inoccuous here as at larger $k$. (ii) It was found in small system 
calculations that $\chi^{CF}$ does not exhaust the
oscillator strength at small wave vectors to the same extent as it does
at larger $k$ \cite {Dev,Wu}. 
This was attributed to the fact that the exciton branch 
comes quite close to the higher bands at small $k$; the exact state mixes 
with higher
band states to reduce its energy, whereas any such mixing is neglected
in the CF approach. (For 2/5 and 3/7, the CF description continues to
work up to the smallest wave vectors available in the finite size
studies.) (iii) In the SMA theory, the $k\rightarrow 0$ 
energy of the 1/3 collective mode was close to twice the 
energy at the minimum.  Our somewhat improved estimate shows that 
the ratio between the two energies is
definitely greater than two, approximately 2.4.  This suggests that the
actual lowest energy excitation at small $k$ may contain {\em two} pairs of
CF-excitons
--- these will have a more complicated quadrupolar structure
\cite {GMP,LZ}. A better understanding of the 
nature of the low-energy excitations in the 
small-$k$ limit would require further investigation.

For other fractions, $\Phi^{UP-CF}$ ane $\chi^{UP-CF}$ 
have a finite fraction of electrons in higher LL's, and the above 
method for the LLL projection is
not useful.  Below, we use a different approach which, though
less accurate, can be applied to the excitons of other FQHE 
states as well.

\section{Unprojected CF exciton: interaction energy}

It has been known that the unprojected CF wave functions 
already reside mostly in the lowest LL \cite {Trivedi}, with very
small kinetic energies per particle. It is natural to ask if they can
be used, without projection, for the computation of various
quantities.  This has been addressed for the charge gap (relevant for
transport), equal to   
the energy required to create a far separated quasielectron-quasihole
pair, which is nothing but an exciton at a large $k$.
It was found that the gap computed with the unprojected wave functions
was smaller than the projected gap by as much as a factor of 2\,-\,3
\cite {unpub}.
This may suggest that the unprojected wave functions are not
relevant for a quantitative understanding.

Fortunately, this turns out not to be the case. 
We will see in this section that the interaction
energy of $\chi^{UP-CF}$ is close to the energy of 
$\chi^{CF}$, up to an overall additive constant, except at
very small $k$. 
In other words, if  the interaction energy of 
$\chi^{UP-CF}$ is called $\Delta V$ (measured relative to 
the unprojected CF ground state), then $\Delta V+\Gamma$ 
provides a good approximation to the exciton dispersion, except at
small $k$. The  constant $\Gamma$ can be fixed by requiring the large
$k$ limit to equal the transport gap (which is experimentally
measurable). The underlying physics will be discussed in the next
section.

We first test this assertion in finite system calculations.
The energy $\Delta V+\Gamma$ (with a suitable choice of $\Gamma$) 
is shown in Figs.~6 and 7 for finite systems, where $\Delta V$ is
computed by variational Monte Carlo.
It indeed captures the essential features of 
$\Delta V_{ex}$; in particular, it obtains correctly the minima and
maxima. It also provides a good quantitative approximation for  
the exciton energy: it is quite accurate for 1/3, and 
reasonably accurate (to within 15-20\%) for 2/5 and 3/7 (for which the
agreement improves with the system size).

The advantages of working with the  unprojected CF wave functions
are that a treatment of large systems becomes possible, and that all 
FQHE states can be considered.  Fig.~4 gives the
dispersion ($\Delta V+\Gamma$, where $\Gamma=0.064
e^2/\epsilon l_{0}$ is obtained from the extrapolation of finite 
system values) for a 20-electron system for the 1/3  FQHE state. 
A lack of any significant size dependence for systems with slightly
larger $N$ shows that these results are close to the thermodynamic 
limit. Let us first concentrate on the range $kl_{0}>0.5$.
Here, $\Delta V+\Gamma$ provides a good approximation for $\Delta V_{p}$.
A deep minimum appears at the expected position 
$kl_{0}\approx 1.4$.  The other two minima at $kl_{0}\approx 2.7$ and 
$kl_{0}\approx 3.5$ are also seen in $\Delta V$, although they
are less pronounced than in $\Delta V_p$. 
$\Delta V$ also underestimates the 
size of the structure in Fig.~7, which seems to
be a general feature.

$\Delta V+\Gamma$ is shown in
Fig.~8 for 1/3, 2/5 and 3/7 for large systems, with 
$\Gamma$ suitably chosen to produce a
reasonable large-$k$ limit. The positions of the two deep
minima for the 2/5 exciton agree well
with those found in the exact diagonalization results of \cite {Su},
as also does the feature that the second minimum is deeper.
The relatively complicated structure in the dispersion 
clarifies why the small system calculations are unable to provide a
coherent picture.  

A curious feature of the exciton dispersion in Fig.~4 is that $\Delta V$ 
bends downward at small wave vectors ($kl_{0}<0.5$ for 1/3). 
The unprojected scheme is not trustworthy here, since 
LL mixing becomes a more serious problem here. (Remember, 
at $L=1$, $\chi^{UP-CF}$ has 
a zero projection on the lowest LL \cite {Dev}).
Indeed, $\Delta V_{p}$ for 1/3 shows no such bending.

\section{Analogy to IQHE}

An insight into the above results can be gained from a 
comparison with the exciton of the $\nu^*=n$ IQHE state
\cite {Kallin}. 

\subsection{positions of minima}

First we consider the number of minima and their positions.
Fig.~9 shows the exciton dispersion of the IQHE states at 
$\nu^*=1$, 2 and 3. These have
one, two and three minima (some of these are relative to a rising
background, and appear as inflection points), respectively,
which corresponds exactly
to the number of strong minima in the exciton dispersions for 1/3, 2/5 and
3/7. To pursue this analogy further, we have shown by arrows the
positions ($L$) of the minima/inflection points of the
corresponding IQHE state in the dispersions of Figs.~2--~7 (in each
case, the IQHE state with the same $N$ has been used to determine the
minima).
For the 1/3 state, this obtains the $L$ of the minimum off by one unit
(which is an insignificant error in the thermodynamic limit)
while for 2/5 and 3/7, it predicts  the positions of minima correctly. 
Given that $L=kl_{0}\sqrt{q}$, and $l_{0}^{-1}\propto \sqrt{B} 
\propto \sqrt{q}$, this
implies that the minima occur at the same $k$ for the FQHE 
and IQHE states related by the CF theory.  

\subsection{Interaction energy of CF exciton}

The exciton contains a quasihole and a quasielectron. 
For sufficiently large $k$, when the quasiparticles are far apart, 
the energy of the CF exciton can be thought of as having
three contributions: the self energy of the quasielectron (which is 
the energy to create an isolated quasielectron), the self energy of  
the quasihole, and the interaction energy between them.
The interaction energy is a monotonic function of the
distance between the quasiparticles, proportional to $1/r$,
when they are far apart.  As they approach one another, there is a 
non-monotonic, oscillatory contribution to
the interaction energy originating from the fact that the 
the quasiparticles are not point-like objects  but have a finite
extent, with oscillatory structure in their density profiles.
At very small $k$, when the size of the exciton becomes of the order
of the size of a single quasiparticle, the distinction between the
self and interaction energies
becomes nebulous, and the above intuitive picture breaks down.

As mentioned earlier, the unprojected CF theory is not able to 
predict the self energies of the quasiparticles 
quantitatively. However, these do not have any $k$ dependence,
and can be corrected by an
additive constant $\Gamma$. The results of the previous section show 
that the unprojected CF approach does a good job of estimating 
the {\em interaction} energy of the quasihole and quasielectron, 
with is responsible for the structure in the exciton dispersion. 

So long as the interaction energy is small compared to the self
energy, the densities of the constituent quasiparticles are not much
affected. The structure in the interaction energy then arises 
largely from the structure in the density profiles of the individual
quasiparticles. 

This motivates us to ask if the unprojected CF theory gives a 
reasonable description of the density profiles of the quasiparticles.
To this end, we first investigate if the FQHE quasiparticles  
have any relation to the IQHE quasiparticles.
The simplest case is of the quasielectron and quasihole of the
$\displaystyle{1\over (2m+1)}$ state. These are related, respectively,
to the quasielectron and quasihole of the $\nu^*=1$ state, i.e., 
to states with the LLL completely occupied
except for a hole, and the LLL completely
occupied plus an additional electron in the second LL. The
associated densities are given by
\begin{equation}
\frac{\rho(r)}{\rho_0}=1-e^{-\frac{1}{2}r^2}\;\;,
\end{equation}
and
\begin{equation}
\frac{\rho(r)}{\rho_0}=1+\frac{1}{2}r^2e^{-\frac{1}{2}r^2}\;\;,
\end{equation}
respectively, where $\rho_0=(2\pi l_0^2)^{-1}$ is the density of the
filled LL. These are plotted in Fig.~10.
The density profiles are different simply because the hole is in the
LLL, while the additional electron resides in the second LL. 
The analogy to the IQHE thus provides a natural explanation of the 
asymmetry between the quasihole and quasielectron. The
charge densities of the quasihole and the quasielectron of the
Laughlin 1/3 state \cite {Morf} are similar to those in Fig.~10.
In particular, the quasielectron of the 1/3 state
has a smoke-ring shape with a dip at the origin.
The maximum of the quasielectron of the $\nu^*=1$ state 
is at $r=\sqrt{2}\: l_{0}$.
This predicts a maximum for the 1/3 quasielectron at $r=\sqrt{6}\:
l_{0}$ (since the $l_{0}$'s at $\nu=1$ and $\nu=1/3$ differ by a
factor of $\sqrt{3}$), which is in excellent agreement with  
$r\approx \sqrt{6}\: l_{0}$ found in earlier studies \cite {Morf}.
The value of $\rho/\rho_0$ at the origin is $1.0$ and that at the maximum is
$1+e^{-1}=1.37$, which should be compared to $\rho/\rho_{0}\approx 
0.93$  and $\rho/\rho_0\approx 1.26$ (where $\rho_0$ now is the
electron density of the 1/3 state),  respectively, for the 1/3
quasielectron (taken from Morf and Halperin \cite {Morf}).
The quasiholes at $\nu=1/3$ and $\nu=1$ also have rather similar
shapes, except near the origin. 
The similarity between the quasiparticles 
of a FQHE state, which are  
completely determined by interactions between electrons confined to
the lowest LL, with the quasiparticles of the corresponding IQHE
state, governed by the LL physics, is quite impressive.
This appearance of the higher-LL-like structure within the lowest LL
is an explicit illustration of the basic principles of the CF theory.
The wave functions of the
quasiparticles of other FQHE states are expected to have more
complicated profiles, since the radial part of the single 
particle wave function in higher LL's has several nodes. 

Multiplication by the Jastrow factor does not change the main features
of the IQHE quasiparticles, although it introduces additional weaker  
oscillations. It was seen in Ref. \cite
{Bonesteel} that $\chi^{UP-CF}$ gives a good description of the
density profile even in the region of overlap between the
quasielectron and quasihole.

\subsection{Small k limit}

Another feature of the unprojected results is that $\Delta V$ 
decreases for small $k$ (apparently vanishing as $k\rightarrow 0$). 
This can again be understood by analogy to the IQHE
collective modes, for which also the interaction energy vanishes in
the limit $k\rightarrow 0$, as guaranteed by the Kohn's theorem \cite
{Kohn}.

The analogy between the FQHE and the IQHE excitons 
is, of course, not true in every detail. In particular, it 
applies only to the strong structure. There are weaker 
minima in the CF exciton dispersion, which have no analog in the
dispersion of the corresponding IQHE exciton.
These appear only after multiplication by the Jastrow 
factor (with or without the LLL projection), and thus refer to
very fine features beyond the mean-field theory. 
These are related to weaker oscillations in 
the density profiles of  the FQHE-quasiparticles that are  not seen 
for the corresponding IQHE-quasiparticles. 
Also, the relative strengths of the strong minima do not
necessarily correspond to those in the IQHE dispersion. For example, at
$\nu^*=2$, the second minimum is much weaker than the first one 
(Fig.~9), unlike at $\nu=2/5$ where the first one is weaker 
(Fig.~8).

The principal minimum for the 1/3 collective mode has been 
interpreted as a precursor of the Wigner crystal instability \cite {GMP}. 
The above interpretation is
cast in terms of the IQHE of composite fermions. A
reconciliation between the two should prove interesting.

\section{Unprojected CF mode: kinetic energy}

Further insight into some features of the above results can 
be gained by considering the kinetic energy of the
various unprojected wave functions. We emphasize that 
it has no direct relevance to the
actual energy of the collective mode (at least in the limit of large
$B$), except that it must be sufficiently small in order for the 
LLL projection to retain the important correlations of the CF theory.

The kinetic energy of the IQHE exciton (measured relative to the
ground state) is precisely $\hbar\omega_{c}^*$.  
We consider here the kinetic energy $\Delta
K$ of the CF exciton, defined as
\begin{equation}
\Delta K = \frac{<\chi^{UP-CF}|K-E_0|\chi^{UP-CF}>}
{<\chi^{UP-CF}|\chi^{UP-CF}>} \;\;,
\end{equation}
where
\begin{equation}
E_{0}=\frac{<\Phi_{n}^{UP-CF}|K|\Phi_{n}^{UP-CF}>}
{<\Phi_{n}^{UP-CF}|\Phi_{n}^{UP-CF}>}\;,
\end{equation}
and $K$ is the kinetic energy operator.
The kinetic energy per electron for the ground state, $E_{0}/N$, is
known to be small \cite {Trivedi}. 

Before coming to numerical calculations, we mention an exact result.
It was proved earlier that the  rigorous upper limit for 
$\Delta K$ for $\nu=1/(2m+1)$ is the cyclotron energy, since 
$\chi^{UP-CF}$ cannot have more than one electron in the second LL.
It has also been shown earlier that at $L=1$, the wave function has
zero projection on to the lowest LL, indicating that it
has at least one electron in the second LL \cite {He,Dev}.
Taken together, one concludes that 
$\chi^{UP-CF}$ has precisely one electron in the
second LL at $L=1$, i.e., its energy approaches the cyclotron energy
in the limit $k\rightarrow 0$.  For other fractions, no exact results
are known.

$\Delta K$ cannot be
evaluated analytically in general.  We have computed it  
by two methods.  For small systems, we expand $\chi^{UP-CF}$ and
$\Phi^{UP-CF}$ in the Slater determinant basis. For 
$\chi^{UP-CF}$, there are a
total of 18,814 basis states for $N=7$ at $\nu=1/3$ 
(in the $L_{z}=0$ subspace,
allowing a maximum of one electron in the second LL),
and 4,448,884 basis states for $N=8$ at $\nu=2/5$ (in the
$L_{z}=0$ subspace, allowing for a maximum of four electrons in the
second and one electron in the third LL).
The amplitudes of various basis states can be obtained by an elaborate
book-keeping.  
Despite the large basis sizes, we are able to 
evaluate $\Delta K$ exactly (numerically) for these systems (Fig.~11), 
since it is not necessary to store the amplitudes of all states.
In these calculations, we put by hand the LL
spacing to be the same for all consecutive LL's.
As expected, $\Delta K$ of the 1/3 state at $L=1$
is exactly equal to the cyclotron energy.  For 2/5 also, 
$\Delta K$ at $L=1$  is close to $\hbar\omega_{c}$
as seen in Fig.~11(b). The actual values for
$\Delta K/\hbar\omega_{c}$ at $L=1$ for 4, 6,
and 8 electrons are 1.0219, 1.0091, and 1.0050, respectively,
approaching unity approximately as $1+0.9N^{-5/2}$.

We have studied bigger systems by Monte Carlo, and the 
results are shown in Fig.~12.  $\Delta K$ is 
small, less than the cyclotron energy in the entire
$k$ range for all FQHE states. 
It approaches $\hbar\omega_{c}$ as $k \rightarrow 0$, in
accordance with the Kohn's theorem (below).
At large $k$, its value is equal to the kinetic energy of a far
separated quasielectron-quasihole pair.
For the 1/3 state, the large-$k$ limit is equal to
0.16 $\hbar\omega_{c}$, which is also the kinetic energy of an
isolated quasielectron \cite {Trivedi} (since the 
quasihole is strictly in the lowest LL in this case). 
For the 2/5 and 3/7 states, $\Delta K$ shows
2 and 3 shallow minima respectively which occur at approximately
the same positions as the minima in $\Delta V^{UP-CF}$, demonstrating
that $\Delta K$ also shares some qualitative 
features of the true exciton dispersion. 
Also note that $\Delta K$ increases at small $k$, which sheds light on
why the results obtained in previous sections become questionable
here.

As discussed earlier, the SMA wave function for the collective 
mode at 1/3 is obtained by CF transformation from the state 
$\rho_{_{k}}\Phi_{1}$. The kinetic energy of this wave function 
is given by
\begin{equation}
\frac{<\Phi_{1}|\rho_{k}^{\dagger}(K-E_{0})\rho_{k}|\Phi_{1}>}{
<\Phi_{1}|\rho_{k}^{\dagger}\rho_{k}|\Phi_{1}>}\;\;.
\end{equation}
Substituting from Eq. (\ref{rhoexp}), this is equal to 
\begin{equation}
\frac{\sum_{m=1}^{\infty} m S(m)}{\sum_{m=1}^{\infty}  S(m)}
\;\hbar\omega_{c}^*\;,
\end{equation}
with $S(m)=2^m m! k^{2m}$,
where we have used the fact that $L_{0}^{m}(x)=1$ for arbitrary $x$.
This yields the kinetic energy of  $\rho_{_{k}}\Phi_{1}$ to be
\begin{equation}
\frac{k^2/2}{1-e^{-k^2/2}}\; \hbar\omega_{c}^*\;.
\label{one}
\end{equation}
It approaches the cyclotron energy in the limit $k\rightarrow
0$, as expected from Kohn's theorem. As $k$ increases,  however,
the kinetic energy of $\rho_{_{k}}\Phi_{1}$ increases, as shown in
Fig.~13 (solid line), indicating an increasing occupation of higher LL's. 
Clearly, it is not an appropriate wave function for the exciton 
of the $\nu=1$ IQHE state. The SMA wave function thus has a large
occupation of higher CF LL's at large $k$.
Fig.~13 also shows (squares) the kinetic energy of 
the `unprojected SMA wave function', 
$\prod_{j<k}(z_{j}-z_{k})^2\rho_{_{k}}\Phi_{1}$, measured relative to
the kinetic energy of the state $\prod_{j<k}(z_{j}-z_{k})^2\Phi_{1}$,
for 20 electrons. It approaches 
$\hbar\omega_{c}$ as $k\rightarrow 0$ for the same reason as 
$\Delta K$ (also see, He {\em et al.} \cite {He}), but increases
rapidly beyond $kl_0\approx 1.5$.

Before closing the Section, we state a generalization of the Kohn's
theorem, which sheds light on the feature seen above that 
the kinetic energy of the unprojected wave functions 
approaches $\hbar\omega_{c}$ in the limit of 
$k\rightarrow 0$ for all FQHE states. 
The Kohn's theorem states 
that for any given {\em eigenstate} $\Psi$, 
with eigenvalue $E$, $a_{0}^{\dagger}\Psi$ is also an 
eigenstate with  eigenvalue $E+\hbar\omega_{c}$, with
\begin{equation}
a_{0}^{\dagger}\equiv \sum_{i=1}^N a_{i}^{\dagger}\;\;,
\end{equation}
where $a_{i}^{\dagger}$ is the LL raising operator.
The unprojected CF wave functions, however, are not
eigenstates of the Hamiltonian (in particular of the kinetic energy).
A weaker form of Kohn's theorem applies to 
a class of wave functions which are not eigenstates of the 
Hamiltonian. More specifically, 
\begin{equation}
\frac{<a_{0}^{\dagger} \Psi|H-E|a_{0}^{\dagger} \Psi>}{<a_{0}^{\dagger} 
\Psi|a_{0}^{\dagger} \Psi>} \;=\; \hbar\omega_c,
\label{Kohn}
\end{equation}
where $H=K+V$ and $E=<\Psi|H|\Psi>/<\Psi|\Psi>$, 
provided that $\Psi$ can be written in a factorized form
\begin{equation}
\Psi=\Phi\; \zeta_{COM}
\end{equation}
where (i) $\Phi$ does not depend on the center-of-mass (COM) coordinates,
(ii) $\zeta_{COM}$ depends {\em only} on the COM coordinates, and
(iii) $\zeta_{COM}$ is an eigenstate of the COM kinetic
energy. All these properties are satisfied by any eigenstate.
This generalized Kohn's theorem can be proven straightforwardly
by transforming to the center of mass and relative coordinates, and
noting that there is no mixing between the COM and relative
coordinates in the kinetic energy $K$, which is a sum of two parts, one
containing only the COM coordinate and the other containing only the
relative coordinates. The operators $a_0^{\dagger}$ and $a_0$
are recognized as the LL raising and lowering operators for the COM
coordinate. The theorem then follows since the application of 
$a_{0}^{\dagger}$ changes the kinetic energy by precisely
$\hbar\omega_c$, but leaves the interaction energy unchanged (since it
is independent of the COM coordinate). 

The unprojected CF wave function for the incompressible FQHE state
at $\nu=n/(2n+1)$ is given by 
\begin{equation}
D^m\Phi'_n \;\exp[-\frac{1}{4}\sum_{i=1}^{N}|z_i|^2]\;\;, 
\end{equation}
where the prime on $\Phi'_n$ denotes that the exponential has
been factored out. To see that it has the same form as the wave
functions in Eq.~(\ref{Kohn}), note that the exponential can be 
expressed as 
\begin{equation}
\exp[-\frac{1}{4}(\sum_{i=1}^{N-1}(z_i-\eta_0)(z_i^*-\eta_0^*)+
|\sum_{i=1}^{N-1}(z_i-\eta_0)
|^2)]\;\; \exp[-\frac{N}{4}\eta_0\eta_0^*]\;\;,
\end{equation}
where $\eta_0=N^{-1}\sum_{j=1}^{N}z_j$ is the COM coordinate.
The factor $D^m\Phi'_n$ is independent of the COM coordinates, since
it is invariant under the replacement $z_{j}\rightarrow z_{j}-\eta_0$).
The COM coordinate thus factors out; in fact,
$\zeta_{COM}=\exp[-\frac{N}{4}\eta_0\eta_0^*]$ is a ground
state of the COM kinetic energy. 

These considerations provide a systematic way of constructing
new trial states at the same energy and also at energies differing 
by $\hbar\omega_c$, by using the angular momentum and LL raising 
operators of the COM coordinate.
This is related to the fact, seen routinely in finite size
studies, that once an eigenenergy appears at any angular momentum
value, it appears at all larger angular momenta as well \cite
{Trugman,Stone}.

\section{Conclusion}

In short, the  following unified picture becomes possible in
the CF framework: the incompressible ground states 
contain an integer number of filled CF-LL's and the excited states  
are obtained by exciting composite fermions to higher CF-LL's.  
Not surprisingly, the same physics provides the key to 
understanding both the ground state and its  
excitations.

We have carried out a detailed study of the neutral excitations of the
1/3, 2/5 and 3/7 states and developed techniques that allow us to
deal with fairly large systems. Predictions have
been made for the rich structure of the collective modes of various
FQHE states. The dispersion of 
the $\displaystyle{n \over (2n+1)}$ state has $n$ strong
minima, and several additional weaker minima. The strong minima can be
explained by analogy to the IQHE, and appear at the same wave vector
as the minima in the collective modes of the IQHE states.
The weaker minima, on the other hand, have  no analog in the FQHE.
The existence of these structures has been corroborated,  whenever
possible, with the help of exact diagonalization study.

There is  qualitative agreement between the results of our 
calculations and those of the Chern-Simons theory. The latter also
obtains flat dispersion at large $k$, several minima, and finite energy 
in the limit $k\rightarrow 0$. A quantitative comparison is not as good,
however. A crucial feature of the CS scheme is that, 
at small $k$, the mode derived from the $n\rightarrow n+1$ 
IQHE mode is pushed up to the cyclotron energy due to the RPA
screening.  This may be related to the feature discovered in Section
VI that $\Delta K \rightarrow \hbar\omega_c$ as $k\rightarrow 0$ for 
the unprojected CF wave function.  A better understanding of these 
and other issues will require further work.

Several effects left out in the above study must
be incorporated before a comparison with experiment may be made. 
Modification in the Coulomb interaction due to the finite 
width of the quantum well \cite{Wellwidth}, LL mixing
\cite {LLmixing}, and 
disorder \cite {Laugha,Disorder} are all known 
to change the numerical values of the 
excitation energies.
We believe that a good first approximation for the experimental 
CM dispersion can be obtained by $\Delta V+\Gamma$, with a choice 
of $\Gamma$ that makes the large-$k$ limit equal to the experimental
transport gap.

This work was supported in part by the National Science Foundation
under Grant no. DMR93-18739. It is a pleasure to acknowledge 
valuable discussions with S.A. Kivelson, A.H. MacDonald, and A. Pinczuk.

\pagebreak

\pagebreak

\begin{figure}
\centerline{\psfig{file=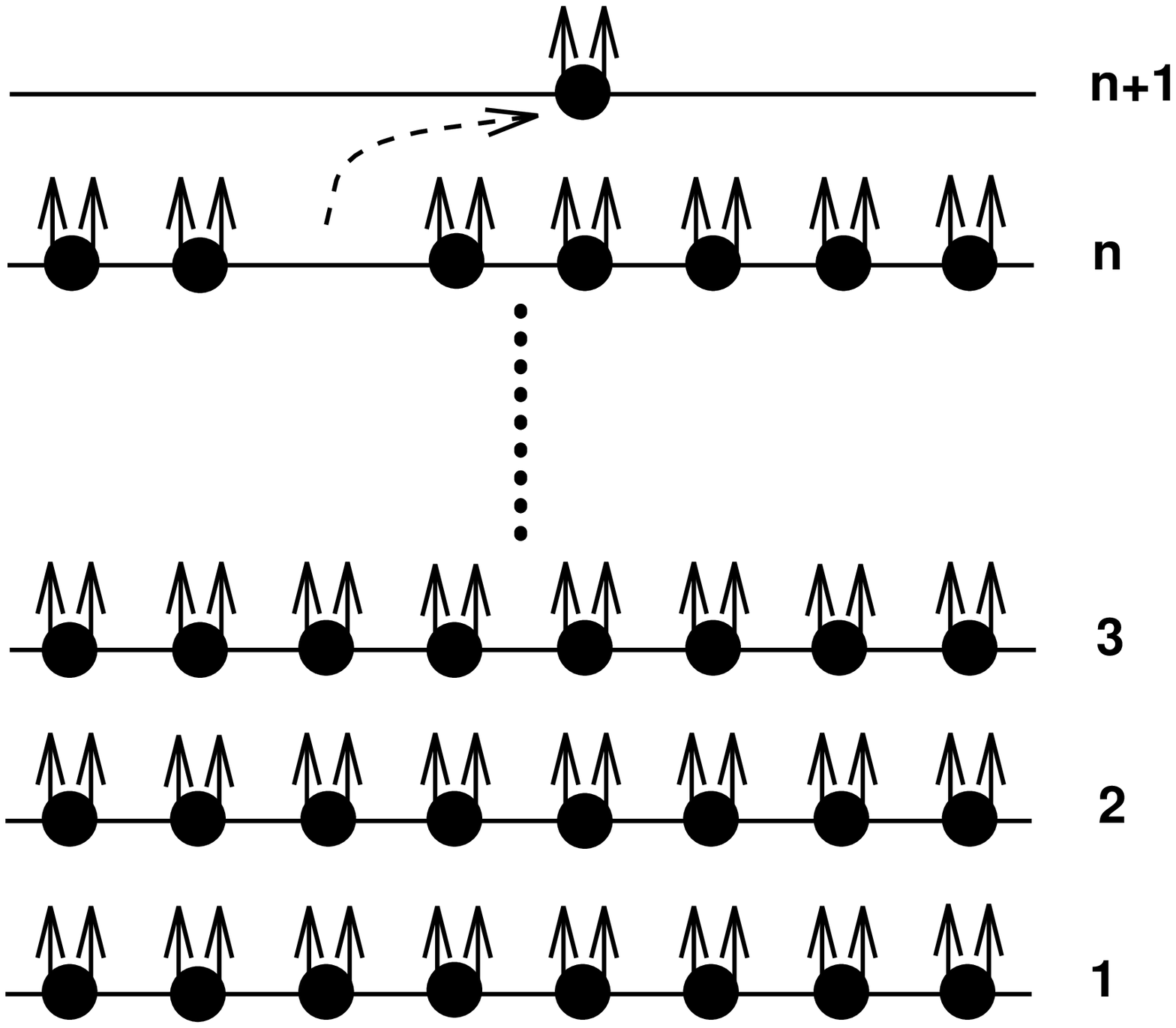,width=5.0in}}
\vspace*{2.0cm}
\caption{Schematic view of a CF exciton. Each filled circle with two
arrows depicts a composite fermion as an electron carrying two flux quanta. 
All composite fermions on a given line are in the same CF Landau
level.  A CF exciton of the $n$ filled CF-LL state
[which corresponds to $\nu=n/(2n+1)$ FQHE state of
electrons] is obtained by promoting a single composite fermion 
from the topmost occupied [$n$th] CF Landau level to the lowest 
unoccupied [$(n+1)$th] CF Landau level. 
}
\label{fig:CF1}
\end{figure}

\begin{figure}
\centerline{\psfig{file=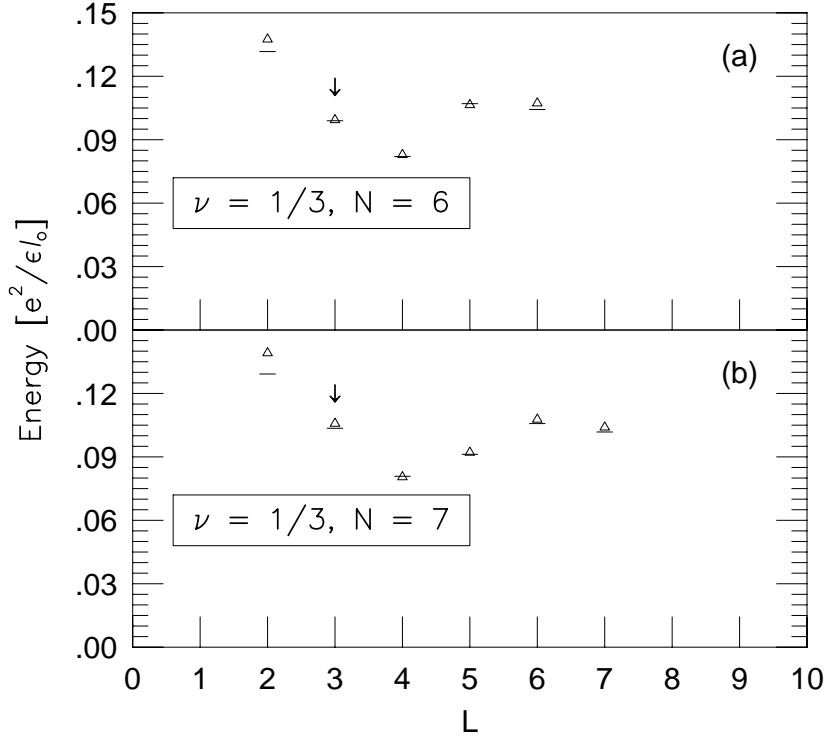,width=5.0in}}
\caption{The exact energies $\Delta V_{ex}$ (dashes) and projected CF
energies $\Delta V_{p}$ (triangles) are shown 
for the exciton of the
1/3 state for systems of 6 and 7 particles.
All interaction energies here and in the following curves are in units of
$e^2/\epsilon l_{0}$, where $l_{0}$ is the magnetic length, and
$\epsilon$ is the background dielectric constant. The arrows 
here and in the following figures 
indicate the angular momenta at which minima or inflection
points occur in the exciton dispersion for the corresponding filled 
LL IQHE state ($\nu=1$ state in this figure) for the same $N$. 
The energy of the ground state at $L=0$ has been set to zero.
}
\label{fig:imp1}
\end{figure}

\begin{figure}
\centerline{\psfig{file=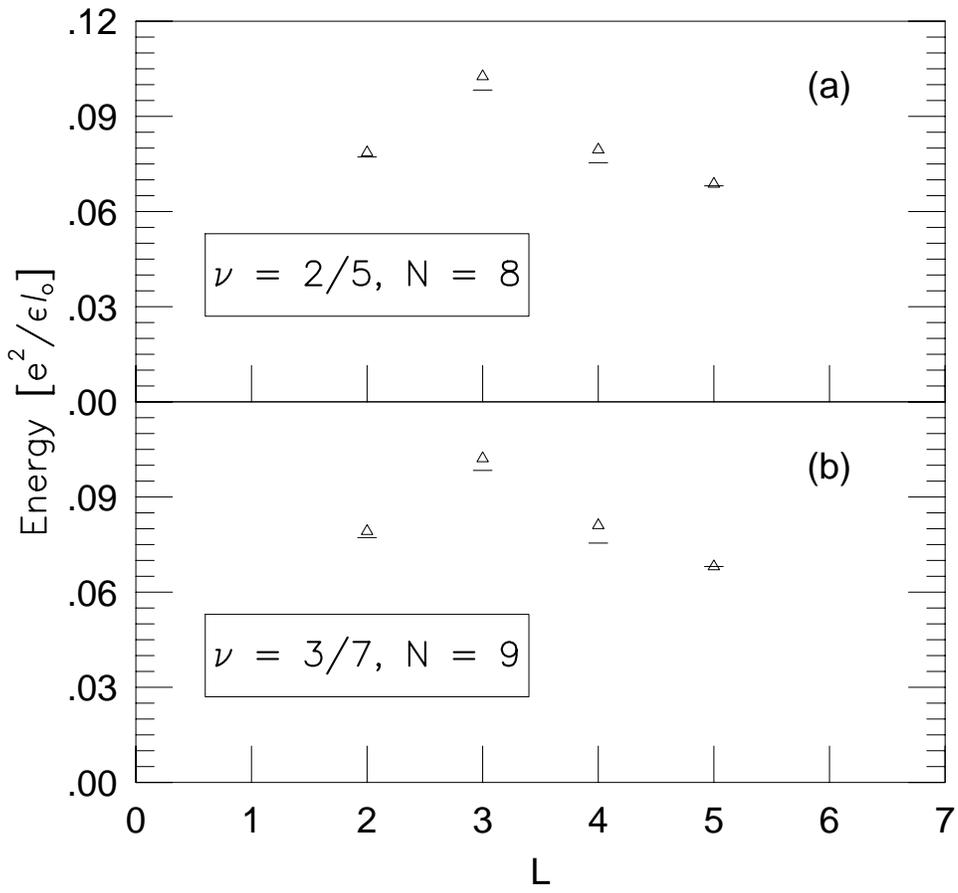,width=6.0in}}
\caption{Same as Fig.~2 for the exciton at $\nu= 2/5$ and 3/7.
}
\label{fig:imp2}
\end{figure}

\begin{figure}
\centerline{\psfig{file=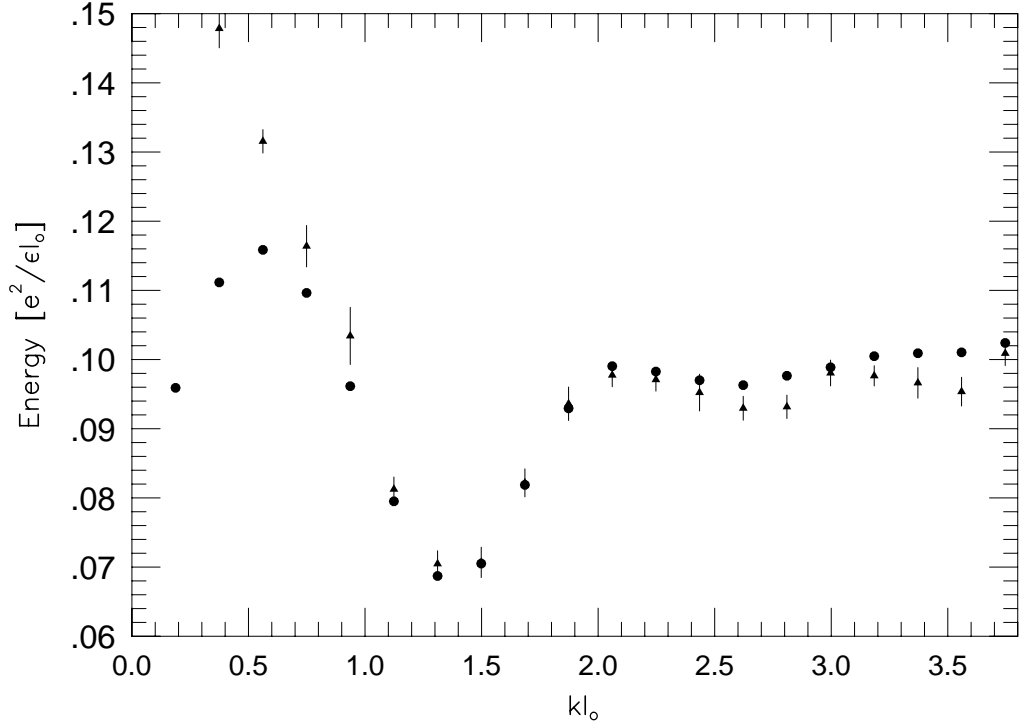,width=6.0in,angle=90}}
\caption{$\Delta V+\Gamma$
with $N=20$ and $\Gamma=0.064 e^2/\epsilon l_{0}$ (filled circles);
and $\Delta V_{p}$ (filled triangles) for 1/3 CF state. $\Delta
V_{p}$ ($\Delta V$) is the interaction energy of the projected 
(unprojected) CF state. The statistical
errors of Monte-Carlo estimation of the energies are shown on
each point for the projected energy; in the case of the unprojected
energies the error is of the order of the size of the filled circles.
}
\label{fig:imp3}
\end{figure}

\begin{figure}
\centerline{\psfig{file=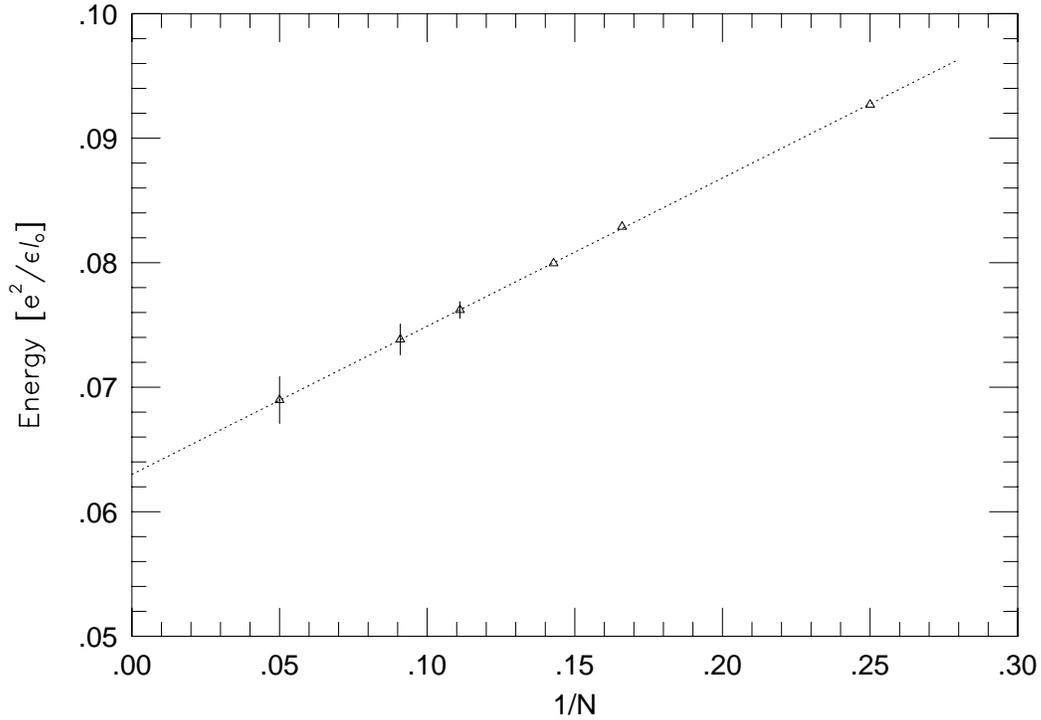,width=6.0in,angle=90}}
\caption{Estimation of $\Delta V_p$ at the minimum of the exciton
dispersion at $\nu=1/3$. For each value of $N$, the minimum was 
determined by smooth polynomial interpolation
in the neighborhood of the minimum to reduce the sensitivity of
our thermodynamic estimation to the discreteness of $L$  
for finite systems. A linear $\chi^2$-fit gives a thermodynamic
estimate of 0.063 $ e^2/\epsilon l_{0}$, with an uncertainty of 2 (3)
in the last digit if four (three) largest systems are used.
}
\label{fig:imp4}
\end{figure}

\begin{figure}
\centerline{\psfig{file=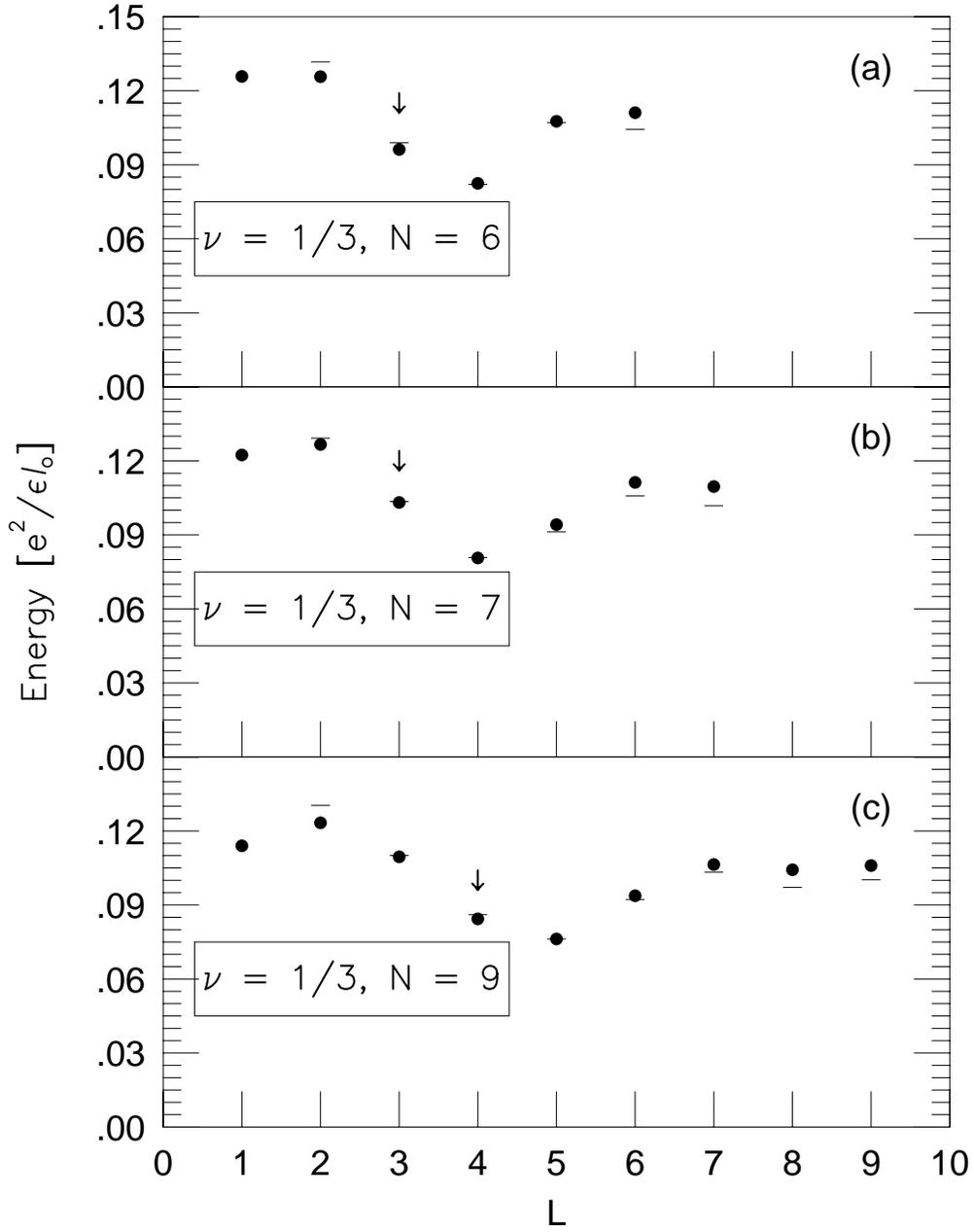,width=6.0in}}
\caption{The exact energies $\Delta V_{ex}$ (dashes)
and $\Delta V+\Gamma$ (filled circles) are shown for the exciton of the
1/3 state for systems of 6, 7 and 9 particles. $\Gamma$ is chosen  so 
as to match the two energies at the minimum.
The exact energies in (c) are taken from Fano {\em et al.} 
\protect\cite {Fano}.}
\label{fig:imp5}
\end{figure}

\begin{figure}
\centerline{\psfig{file=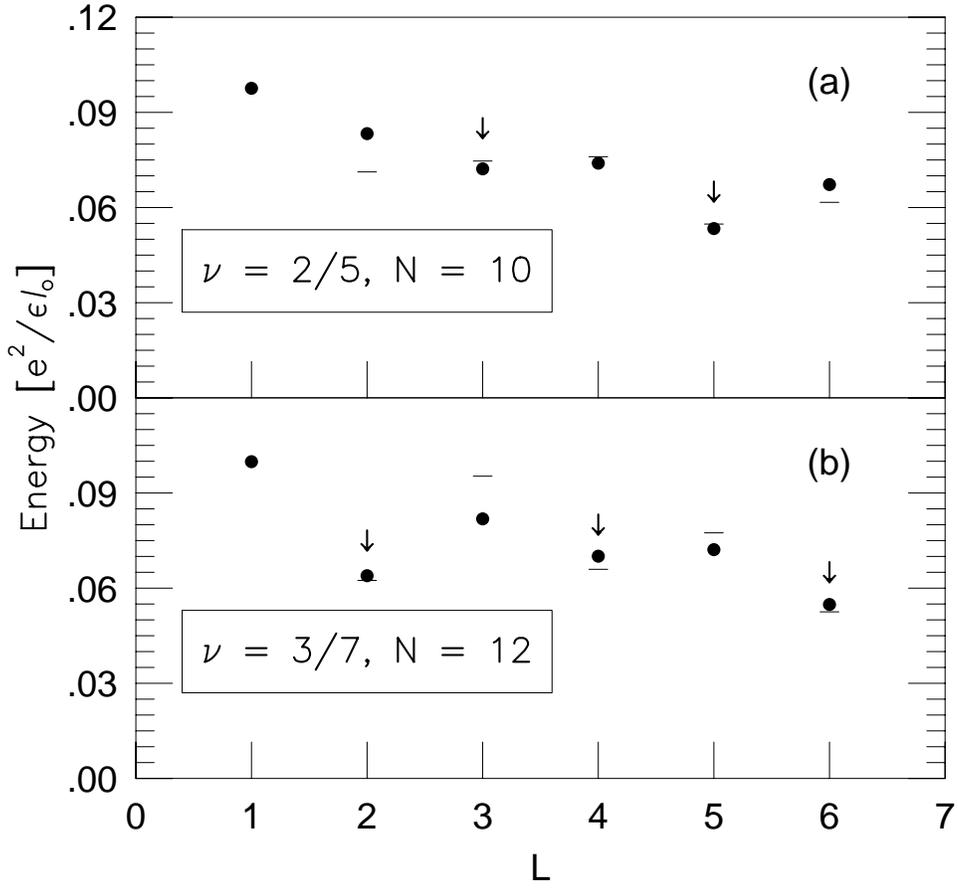,width=6.0in}}
\caption{Same as Fig.~4 for the 2/5 and 3/7 state.  The
exact energies in (a) are taken from d'Ambrumenil and Morf
\protect\cite {Ambrumenil} and in (b) from He  {\em et
al.} \protect\cite {Lopez2}.
}
\label{fig:imp6}
\end{figure}

\begin{figure}
\centerline{\psfig{file=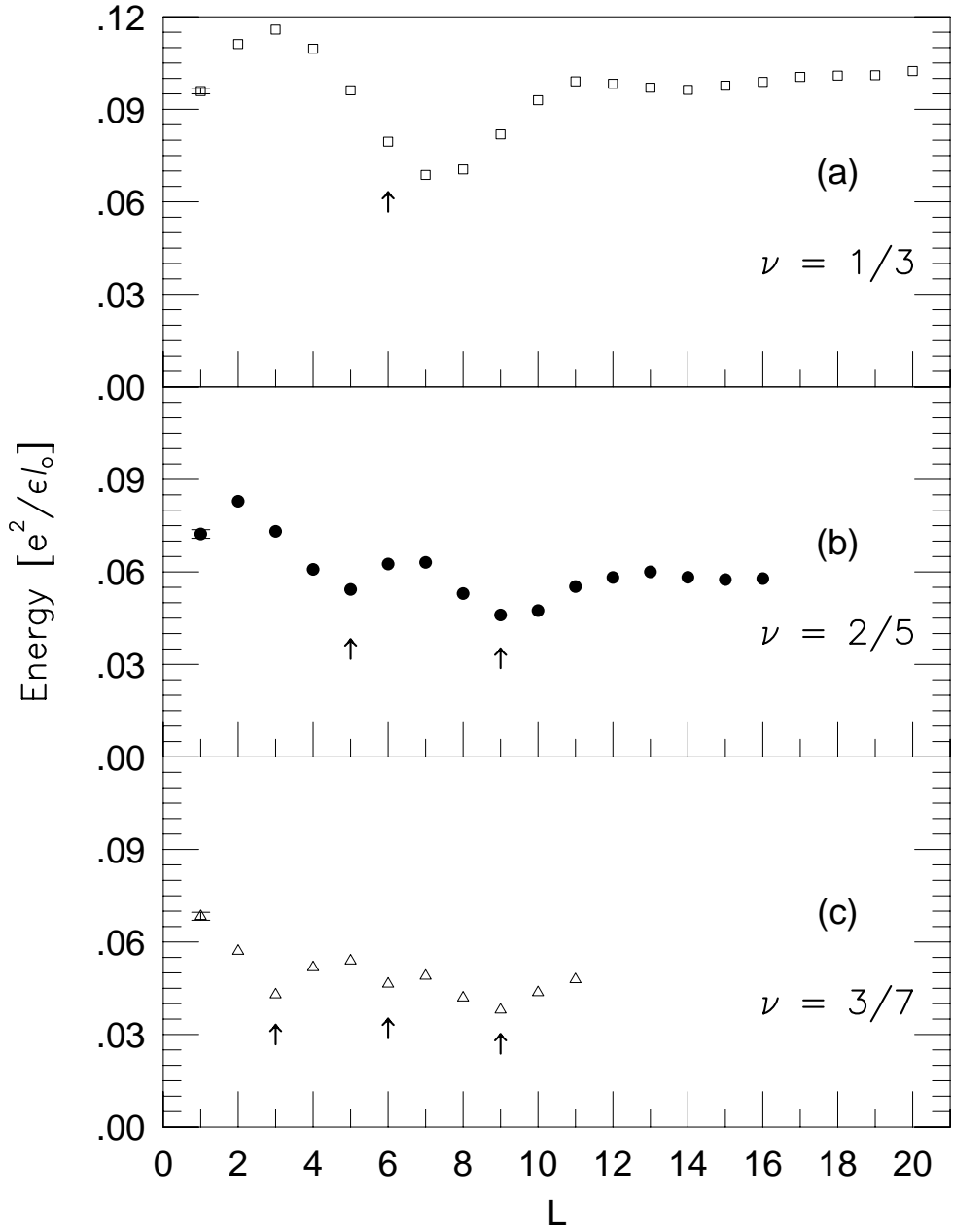,width=6.0in}}
\caption{$\Delta V+\Gamma$ is shown for (a) $\nu=1/3$,
$N=20$, $\Gamma=0.064 e^2/\epsilon l_{0}$; (b) $\nu=2/5$, 
$N=30$, $\Gamma=0.037 e^2/\epsilon l_{0}$; and (c) $\nu=3/7$, 
$N=27$, $\Gamma=0.028  e^2/\epsilon l_{0}$. The typical error bar
is shown on the first point in each case. 
}
\label{fig:imp7}
\end{figure}

\begin{figure}
\centerline{\psfig{file=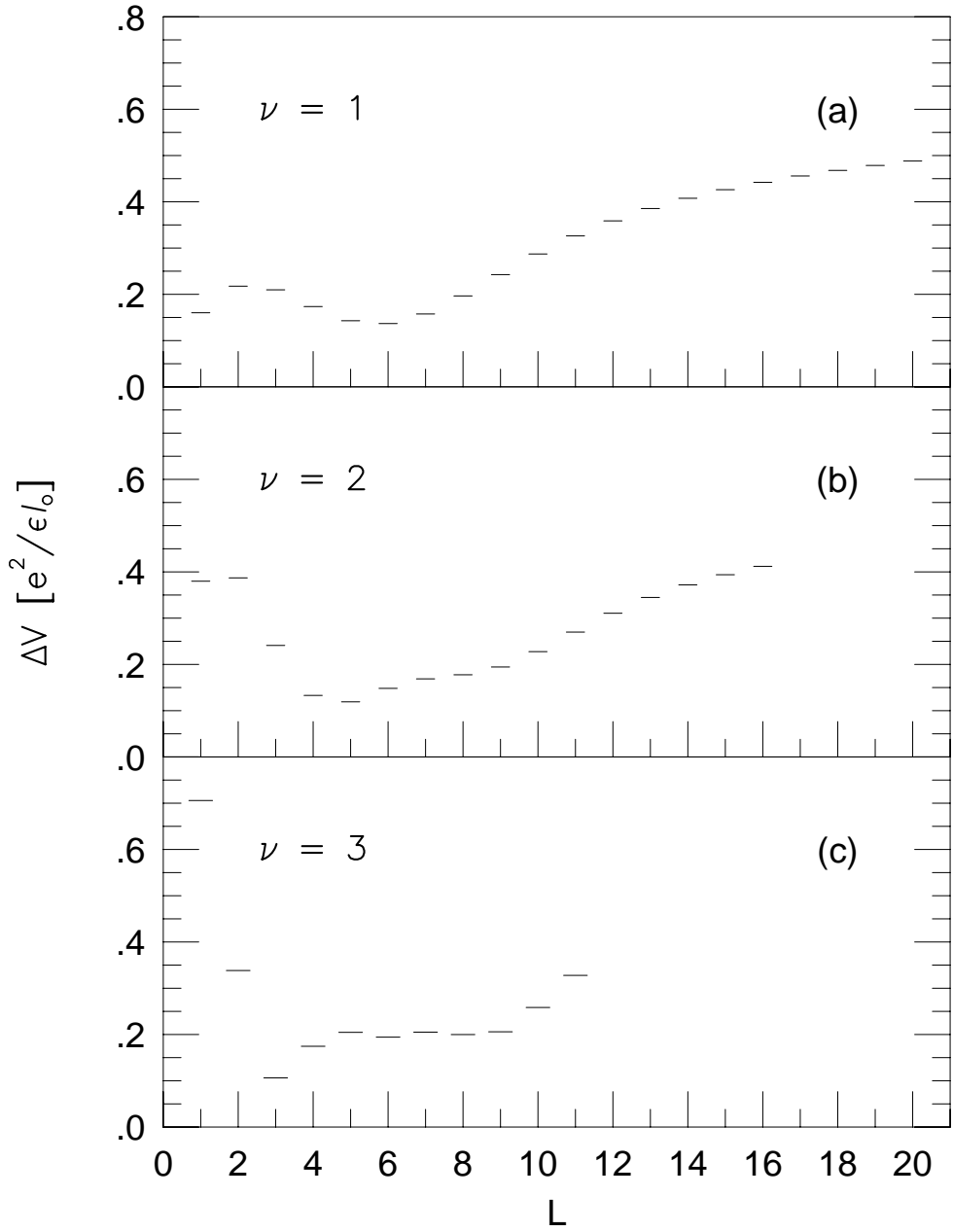,width=6.0in}}
\caption{$\Delta V$ for lowest energy
exciton of (a) $\nu = 1$,
$N = 20$, (b) $\nu = 2$, $N = 30$, and (c) $\nu = 3$, $N = 27$.
These correspond to the FQHE systems in Fig.~8.
There are $n$ minima/inflection points in the lowest branch of the
collective mode of the $\nu = n$ IQHE state.
}
\label{fig:imp8}
\end{figure}

\begin{figure}
\centerline{\psfig{file=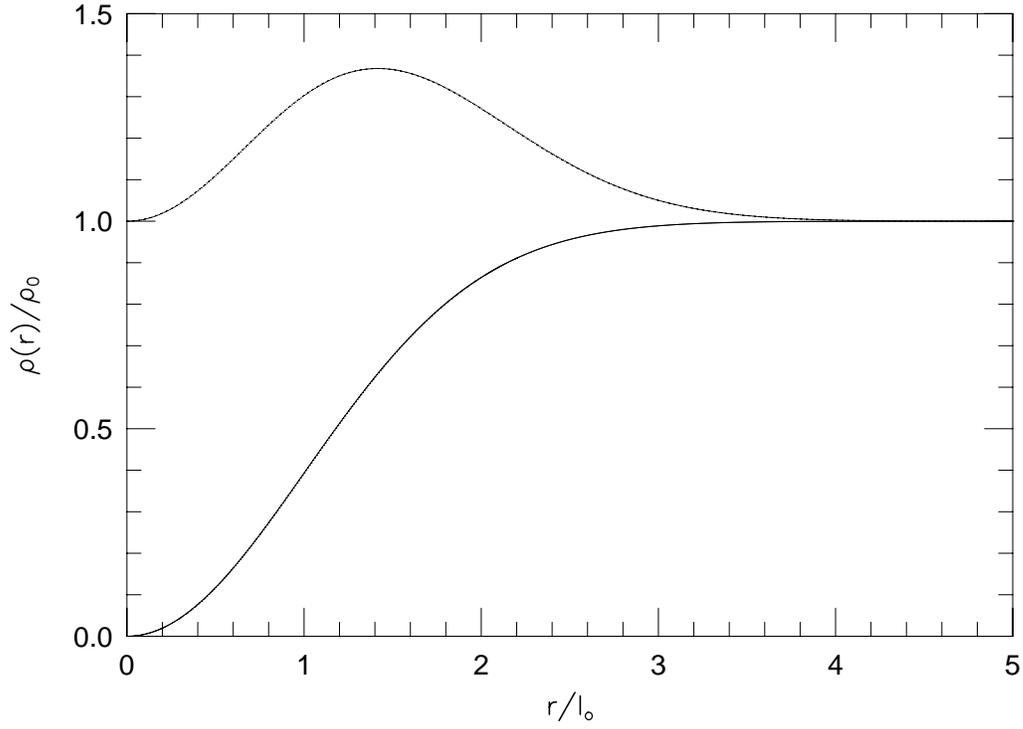,width=6.0in,angle=90}}
\caption{The charge density of the (a) $\nu=1$ quasihole (solid line), 
which is the fully occupied lowest LL except for a hole at the origin, 
and (b) $\nu=1$ quasielectron (dashed line), which is the fully 
occupied lowest LL plus an electron in the second LL at the origin.
}
\label{fig:imp9}
\end{figure}

\begin{figure}
\centerline{\psfig{file=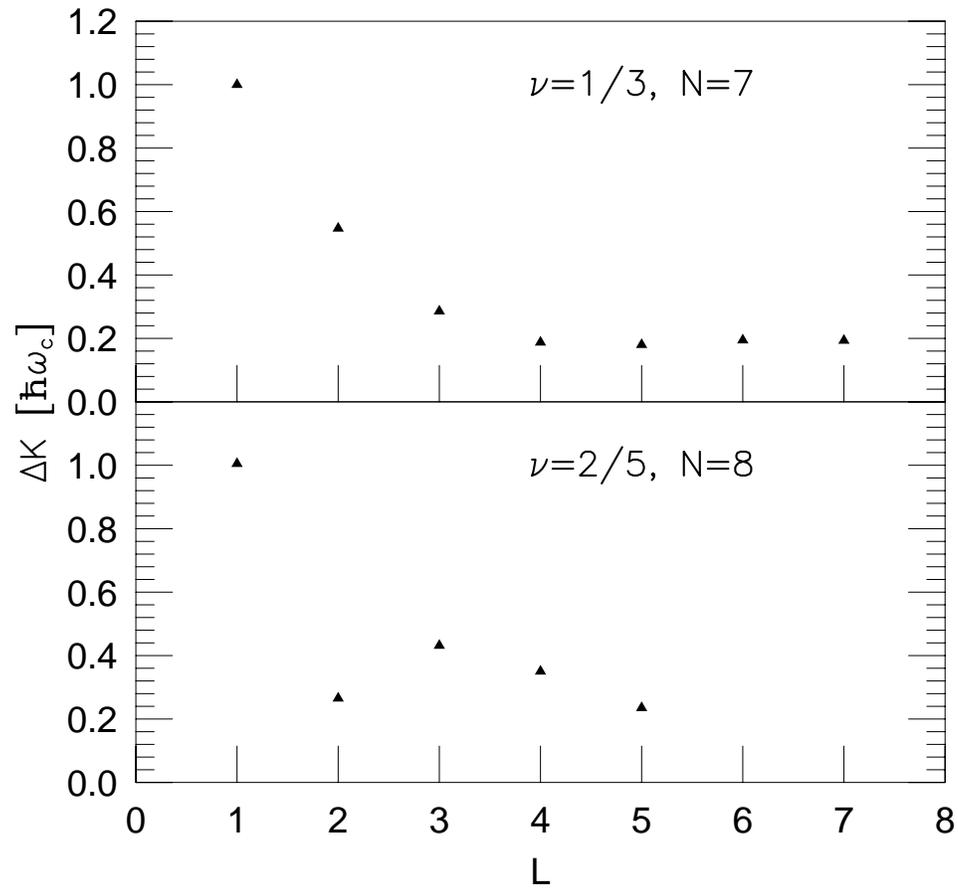,width=6.0in}}
\caption{Exact $\Delta K$ for (a) $\nu=1/3$, $N=7$; 
(b) $\nu=2/5$, $N=8$. 
}
\label{fig:imp10}
\end{figure}

\begin{figure}
\centerline{\psfig{file=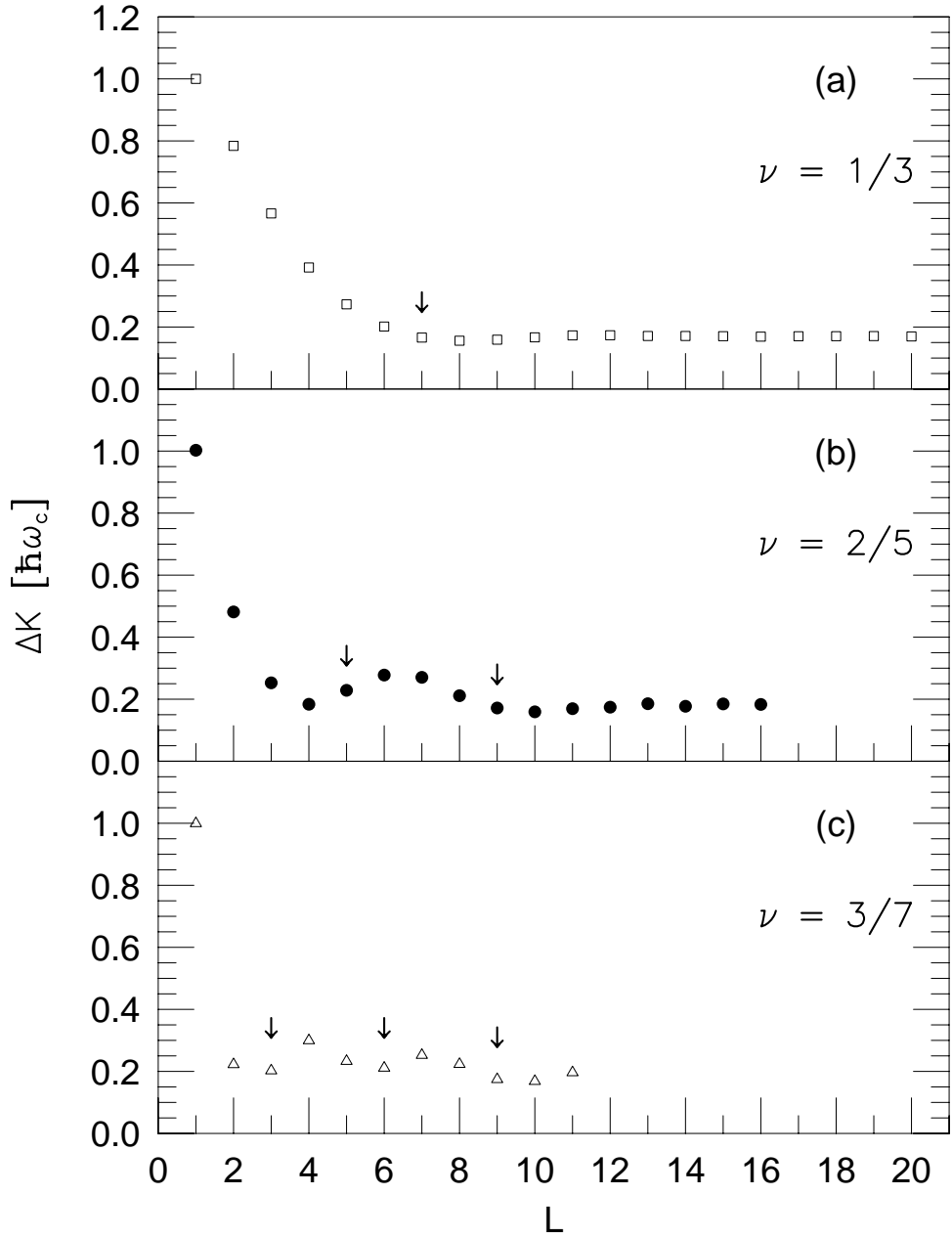,width=6.0in}}
\caption{Monte Carlo estimate for $\Delta K$ using 
the unprojected CF exciton  
wave function for (a) 1/3 state of 20 particles,
(b) 2/5 state of 30 particles, and (c) 3/7 state of 27
particles. The statistical error is smaller 
than the size of the symbol used. 
}
\label{fig:imp11}
\end{figure}

\begin{figure}
\centerline{\psfig{file=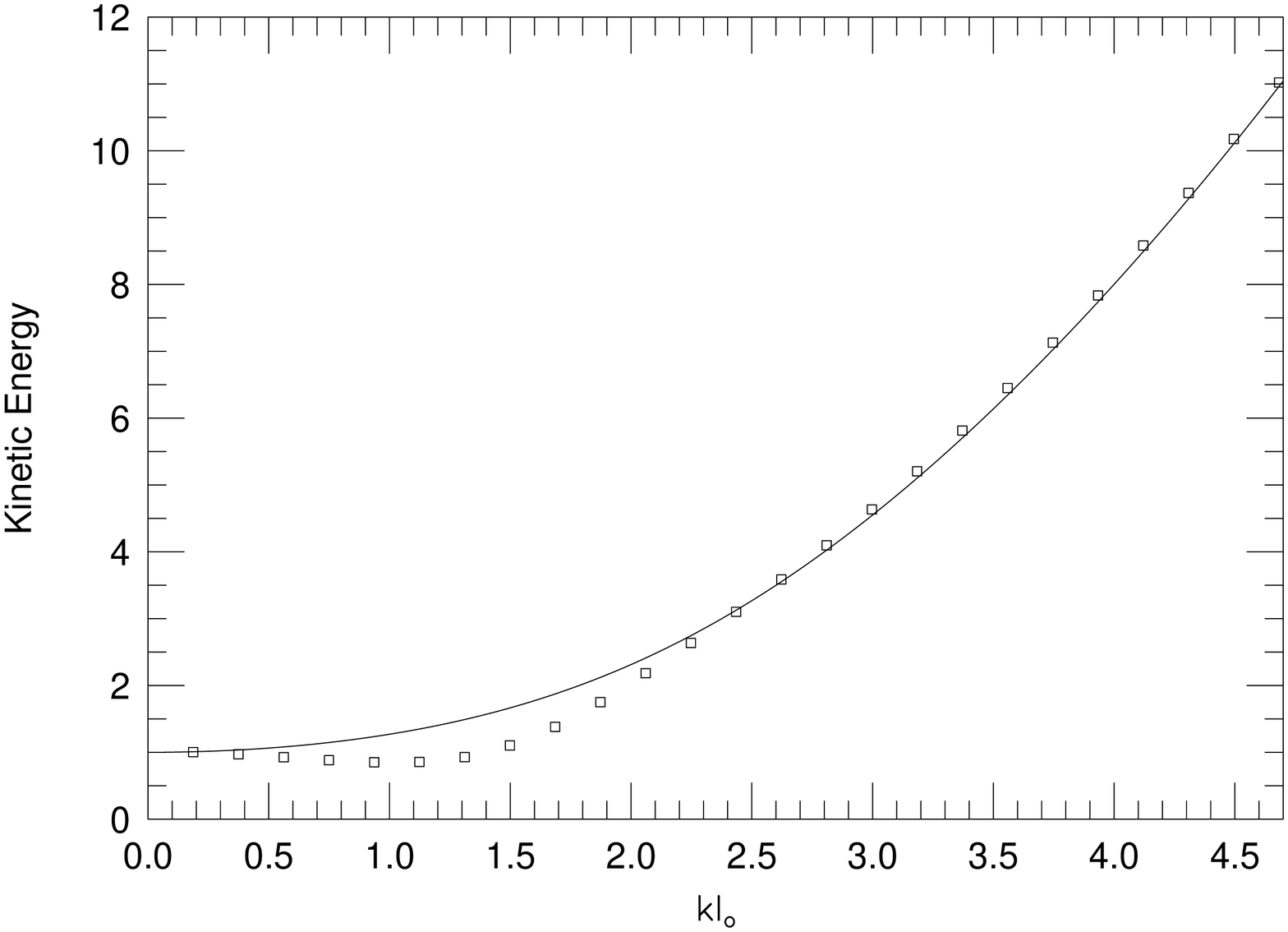,width=6.0in,angle=90}}
\caption{The kinetic energy of the `unprojected
SMA state', in units of $\hbar\omega_{c}$, for the collective mode of the  
1/3 state ($N=20$), shown by squares.
The solid line plots Eq.~(\protect\ref{one}) in unit
of $\hbar\omega_{c}^*$.
}
\label{fig:imp12}
\end{figure}


\begin{thebibliography}{99}



\bibitem{KvonK} K. von Klitzing, G. Dorda and M. Pepper, Phys. Rev.
Lett. {\bf 45}, 494 (1980); D.C. Tsui, H.L. Stormer and A.C. Gossard,
Phys. Rev. Lett. {\bf 48}, 1559 (1982).

\bibitem{Laughlin1} R.B. Laughlin, Phys. Rev. Lett. {\bf 50}, 1395 (1983).

\bibitem{GMP} S.M. Girvin, A.H. MacDonald and P.M. Platzman, Phys.
Rev. Lett. {\bf 54} 581, 1985; Phys. Rev. B {\bf 33}, 2481 (1986).

\bibitem{Feynman} {\em Statistical Mechanics} by R.P. Feynman
(Benjamin Reading, Mass. 1972).

\bibitem{Laugha} R.B. Laughlin, Physica {\bf 126 B}, 254 (1985).

\bibitem{Su} W.P. Su and Y.K. Wu, Phys. Rev. B {\bf 36}, 7565
(19987).  

\bibitem {He} S. He, S.H. Simon
and B.I. Halperin, Phys. Rev. B {\bf 50}, 1823 (1994).

\bibitem{Pinczuk1} A. Pinczuk {\em et al.},
Phys. Rev. Lett. {\bf 61}, 2701 (1988).

\bibitem{Pinczuk2} L.L. Sohn {\em et al.}, Solid State Commun. {\bf
93}, 897 (1995).

\bibitem{Pinczuk} A. Pinczuk {\em et al.},  
Phys. Rev. Lett. {\bf 70} 3983, 1993; Semiconductor
Science and Technology, vol. {\bf 9}, 1865 (1994).

\bibitem{Mellor}C.J. Mellor {\em et al.}, 
Phys. Rev. Lett. {\bf 74}, 2339 (1995).

\bibitem{Jain} J.K. Jain, Phys. Rev. Lett. {\bf 63}, 199 (1989);
Phys. Rev. B {\bf 41}, 7653 (1990); Science {\bf 266}, 1199 (1994).

\bibitem{Dev} G. Dev and J.K. Jain, Phys. Rev. Lett. {\bf 69}, 2843
(1992). 

\bibitem{Wu} X.G. Wu and J.K. Jain, Phys. Rev. B {\bf 51}, 1752
(1995).

\bibitem{Goldhaber} A.S. Goldhaber and J.K. Jain, Phys. Lett. A 
{\bf 199}, 267 (1995).

\bibitem{Bonesteel} N.E. Bonesteel, Phys. Rev. B {\bf 51}, 9917 (1995).

\bibitem{unpub} X.G. Wu, R.K. Kamilla, and J.K. Jain,  unpublished.

\bibitem{Lopez1} 
A. Lopez and E. Fradkin, Phys. Rev. B {\bf 44}, 5246 (1991).

\bibitem{Lopez2} 
A. Lopez and E. Fradkin, Phys. Rev. B {\bf 47}, 7080 (1993);
S.H. Simon and B.I. Halperin, Phys. Rev. B {\bf 48},
17368 (1993); {\em ibid}, {\bf 50},1807 (1994); S. He, S.H. Simon
and B.I. Halperin, Phys. Rev. B {\bf 50}, 1823 (1994); X.C. Xie, 
{\em ibid.}, {\bf 49}, 16833 (1994); L. Zhang, {\em ibid.}, {\bf 51},
4645 (1995); and Phys. Rev. B, in print (SISSA preprint no. 
cond-mat/9506113).

\bibitem{Binder} {\em Monte Carlo Methods in Statistical Physics},
edited by K. Binder (Springer-Verlag, NY 1979), and references
therein; Monte Carlo Simulations of Disordered Systems, by
S. Jain (World Scientific, Singapore 1992).

\bibitem{Ceperley} D. Ceperley, G.V. Chester, M.H. Kalos; Phys. Rev. B
{\bf 16}, 3081 (1977); S. Fahy, X.W. Wang, and S.G. Louie,
Phys. Rev. B {\bf 42}, 3503 (1990).

\bibitem{KWJ} R.K. Kamilla, X.G. Wu, and J.K. Jain, Phys. Rev. Lett.
{\bf 76}, 1332 (1996).

\bibitem{Trivedi} N. Trivedi and J.K. Jain, Mod. Phys. Lett. B {\bf
5}, 503 (1991).

\bibitem {Kallin} C. Kallin and B.I. Halperin, Phys. Rev. B {\bf 30},
5655 (1984); and the references therein.

\bibitem{CNYang} T.T. Wu and C.N. Yang, Nucl. Phys. B {\bf 107}, 365
(1976); T.T. Wu and C.N. Yang, Phys. Rev. D {\bf 16}, 1018 (1977).

\bibitem{book} See, F.D.M. Haldane in {\em The Quantum Hall Effect},
edited by R.E. Prange and S.M. Girvin (Springer-Verlag, NY 1990).

\bibitem{Fano} G. Fano, F. Ortolani and E. Colombo, Phys. Rev. B 
{\bf 34}, 2670 (1986).

\bibitem{LZ} D.H. Lee and S.C. Zhang, Phys. Rev. Lett. {\bf 66}, 1220
(1991).

\bibitem{Ambrumenil} N. d'Ambrumenil and R. Morf, Phys. Rev. B
{\bf 40}, 6108 (1989).

\bibitem{Morf} E.g., see R. Morf and B.I. Halperin, Phys. Rev. B {\bf 33},
2221 (1986).

\bibitem{Kohn} W. Kohn, Phys. Rev. B {\bf 123}, 1242 (1961).

\bibitem{Trugman} S.A. Trugman and S. Kivelson, Phys. Rev. B
{\bf 31}, 5280 (1985).

\bibitem{Stone} M. Stone, H.W. Wyld, and R.L. Schult, Phys. Rev. B
{\bf 45}, 14156 (1992).

\bibitem{Wellwidth} F.C. Zhang and S. Das Sarma, Phys. Rev. B {\bf
33},
2903 (1986); D. Yoshioka, J. Phys. Soc. Jpn. {\bf 55},
885 (1986).

\bibitem{LLmixing} D. Yoshioka, J. Phys. Soc. Jpn. {\bf 53},
3740 (1984); X. Zhu and S.G. Louie, Phys. Rev. 
Lett. {\bf 70}, 339 (1993)
V. Melik-Alaverdian and N.E. Bonesteel, Phys. Rev. B {\bf 52}, R17032
(1996); Rodney Price and S. Das Sarma, preprint.

\bibitem{Disorder} A.H. MacDonald, A.H. Liu, S.M. Girvin and
P.M. Platzman, Phys. Rev. B {\bf 33}, 4014 (1986), S.M. Girvin
in {\em The Quantum Hall Effect}, edited by R.E. Prange and 
S.M. Girvin (Second Edition, Springer-Verlag, NY 1990).

\end{thebibliography}
\end{document}